\documentclass[iop]{emulateapj}
\usepackage{apjfonts}
\usepackage{graphicx}
\usepackage{color}
\usepackage{amsmath}
\newcommand{\noprint}[1]{}

\newcommand{\figsetend}{}
\newcommand{\figsetgrpstart}{}
\newcommand{\figsetgrpend}{}

\newcommand{\figsetgrpnum}[1]{\noprint{#1}}
\newcommand{\figsetgrptitle}[1]{\noprint{#1}}
\newcommand{\figsetplot}[1]{\noprint{#1}}
\newcommand{\figsetgrpnote}[1]{\noprint{#1}}
\def\d3{$\delta_{3}$ }
\def\1d3{$(1 + \delta_{3})$ }
\def\l1d3{$\log_{10}(1 + \delta_{3})$ }

\def\s3{$\Sigma_{3}$}
\def\ha{H$\alpha$}
\def\hb{H$\beta$}

\def\othree{[OIII] 5007}
\def\ntwo{[NII] 6584}

\def\24m{24 $\mu$m}

\def\sm{$\rm~M_{*}$}

\def\kms{${\rm km~s^{-1}}$ }

\def\Msolar{$\rm M_{\odot}$}

\def\rmxaa{RMxAA}

\def\sigsm{$\Sigma_{*}$}
\def\sigsfr{$\Sigma_{\rm SFR}$}

\def\hi{$\rm H\textsc{i}$}
\def\h2{$\rm H_{2}$}
\def\Mh2{$\rm M_{H_{2}}$}
\def\MHI{$\rm M_{H\textsc{i}}$}
\def\sigh2{$\Sigma_{\rm H_{2}}$}
\def\fgas{$f_{\rm gas}$}
\def\fh2{$f_{\rm H_{2}}$}

\def\Re{$R_{e}$}
\def\co{$^{12}$CO(1-0)}
\def\coprime{$^{13}$CO(1-0)}

\shorttitle{ALMaQUEST Survey}
\shortauthors{Lin et al.}

\begin{document}

%\title{The ALMaQUEST Survey II: The Origin of Star-Forming Main Sequence with ALMA and MaNGA (or the Scaling Relation between Star Formation Rate, Gas Mass, and Stellar Mass)}

\title{ALM\lowercase{a}QUEST - IV. The ALMA-M\lowercase{a}NGA QUEnching and STar formation (ALM\lowercase{a}QUEST) Survey}

%\title{The ALMA-MaNGA QUEnching and STar formation (ALMaQUEST) Survey II: The molecular gas main sequence and the origin of the star forming main sequence}

\author{Lihwai Lin \altaffilmark{1}, Sara L. Ellison \altaffilmark{2}, Hsi-An Pan \altaffilmark{3}, Mallory D. Thorp \altaffilmark{2}, Yung-Chau Su \altaffilmark{1,4}, Sebasti\'{a}n F. S\'{a}nchez \altaffilmark{5}, Francesco Belfiore \altaffilmark{6,7}, M. S. Bothwell \altaffilmark{8,9},  Kevin Bundy \altaffilmark{10}, Yan-Mei Chen \altaffilmark{11,12}, Alice Concas\altaffilmark{8,9}, Bau-Ching Hsieh \altaffilmark{1}, Pei-Ying Hsieh \altaffilmark{1}, Cheng Li \altaffilmark{13}, Roberto Maiolino \altaffilmark{8,9}, Karen Masters \altaffilmark{14}, Jeffrey A. Newman\altaffilmark{15,16}, Kate Rowlands \altaffilmark{17,18}, Yong Shi \altaffilmark{19}, Rebecca Smethurst \altaffilmark{20},  David V. Stark \altaffilmark{14}, Ting Xiao\altaffilmark{21},  Po-Chieh Yu \altaffilmark{22}}

\altaffiltext{1}{Institute of Astronomy \& Astrophysics, Academia Sinica, Taipei 10617, Taiwan; Email: lihwailin@asiaa.sinica.edu.tw}%lihwai, 
\altaffiltext{2}{Department of Physics \& Astronomy, University of Victoria, Finnerty Road, Victoria, British Columbia, V8P 1A1, Canada}%sara,mallory
\altaffiltext{3}{Max-Planck-Institut f\"ur Astronomie, K\"onigstuhl 17, D-69117 Heidelberg, Germany}%hsi-an
\altaffiltext{4}{Department of Physics, National Taiwan University, 10617, Taipei, Taiwan}%jing-hua, chin-hao, yungchau
\altaffiltext{5}{Instituto de Astronom\'ia, Universidad Nacional Aut\'onoma de  M\'exico, Circuito Exterior, Ciudad Universitaria, Ciudad de M\'exico 04510, Mexico}% sanchez
\altaffiltext{6}{European Southern Observatory, Karl-Schwarzschild-Str. 2, Garching bei M$\ddot{u}$nchen, 85748, Germany}%fracesco
\altaffiltext{7}{INAF -- Osservatorio Astrofisico di Arcetri, Largo E. Fermi 5, I-50157, Firenze, Italy}
\altaffiltext{8}{Cavendish Laboratory, University of Cambridge, 19 J. J. Thomson Avenue, Cambridge CB3 0HE, United Kingdom}%roberto, matt
\altaffiltext{9}{University of Cambridge, Kavli Institute for Cosmology, Cambridge, CB3 0HE, UK.}%roberto, matt
\altaffiltext{10}{UCO/Lick Observatory, University of California, Santa Cruz, 1156 High St. Santa Cruz, CA 95064, USA}%kevin bundy
\altaffiltext{11}{Department of Astronomy, Nanjing University, Nanjing 210093, China}%yan-mei
\altaffiltext{12} {Key Laboratory of Modern Astronomy and Astrophysics (Nanjing University), Ministry of Education, Nanjing 210093, People's Republic of China}%yong,yanmei
\altaffiltext{13} {Tsinghua Center for Astrophysics and Physics Department, Tsinghua University, Beijing 100084, China}%cheng
\altaffiltext{14} {Department of Physics and Astronomy, Haverford College, 370 Lancaster Ave, Haverford, PA 19041, USA}%karen
\altaffiltext{15}{Department of Physics and Astronomy, University of Pittsburgh, 3941 O'Hara Street, Pittsburgh, PA 15260, USA}%jeff
\altaffiltext{16}{Pittsburgh Particle physics, Astrophysics, and Cosmology Center (PITT PACC)}%jeff
%%\altaffiltext{15} {Indian Institute of Astrophysics, II Block, Koramangala, Bengaluru 560 034, INDIA}%ramya
\altaffiltext{17} {Space Telescope Science Institute, 3700 San Martin Dr Baltimore, MD 21218, USA}%kate
\altaffiltext{18} {Department of Physics \& Astronomy, Johns Hopkins University, Bloomberg centre, 3400 N. Charles St., Baltimore, MD 21218, USA}%kate
\altaffiltext{19} {School of Astronomy and Space Science, Nanjing University, Nanjing 210093, China}%yong
\altaffiltext{20} {Oxford Astrophysics, Department of Physics, University of Oxford, Denys Wilkinson Building, Keble Road, Oxford, OX1 3RH, UK}%rebecca
\altaffiltext{21}{Department of Physics, Zhejiang University, Hangzhou, Zhejiang 310027, China}%ting xiao
\altaffiltext{22}{Yuan Ze University College of General Studies}%po-chieh

\begin{abstract}
The ALMaQUEST (ALMA-MaNGA QUEnching and STar formation) survey is a program with spatially-resolved $^{12}$CO(1-0) measurements obtained with the Atacama Large Millimeter Array (ALMA) for 46 galaxies selected from the Mapping Nearby Galaxies at Apache Point Observatory (MaNGA) DR15 optical integral-field spectroscopic survey. The aim of the ALMaQUEST survey is to investigate the dependence of star formation activity on the cold molecular gas content at kpc scales in nearby galaxies.  The sample consists of galaxies spanning a wide range in specific star formation rate (sSFR), including starburst (SB), main-sequence (MS), and green valley (GV) galaxies. In this paper, we present the sample selection and characteristics of the ALMA observations, and showcase some of the key results enabled by the combination of spatially-matched stellar populations and gas measurements. Considering the global (aperture-matched) stellar mass, molecular gas mass, and star formation rate of the sample, we find that the sSFR depends on both the star formation efficiency (SFE) and the molecular gas fraction ($f_{\rm H_{2}}$), although the correlation with the latter is slightly weaker. Furthermore, the dependence of sSFR on the molecular gas content (SFE or $f_{\rm H_{2}}$) is stronger than that on either the atomic gas fraction or the molecular-to-atomic gas fraction, albeit with the small $\rm H\textsc{i}$ sample size. On kpc scales, the variations in both SFE and $f_{\rm H_{2}}$ within individual galaxies can be as large as 1-2 dex thereby demonstrating that the availability of spatially-resolved observations is essential to understand the details of both star formation and quenching processes.

\end{abstract}


\section{INTRODUCTION}

Star-forming galaxies are known to form a tight relationship in the total star formation rate (SFR) -- total stellar mass (\sm) plane, dubbed the global star forming main sequence \citep[SFMS or MS;][]{bri04,noe07,elb07,dad07,whi12}. The normalization, shape, and scatter of the SFMS are found to depend on environment, morphology, and the redshift of galaxies \citep[e.g.,][]{lin12,whi12,koy13,lin14,spe14,jia18}. As star formation is regulated by both the amount of available gas and internal feedback processes, mapping the molecular gas, which is the fuel for star formation, is therefore key to understanding how galaxies evolve within and across the global SFMS.  Likewise, the molecular gas content holds important clues as to why galaxies eventually halt their star formation and become quiescent.

The COLD GASS survey \citep[CO Legacy Database for GASS,][]{sai11} was the first systematic CO survey for hundreds of local galaxies with \sm~$> 10^{10}$\Msolar. The `extended' COLD GASS \citep[xCOLD GASS,][]{sai17} pushed the sample down to \sm~$> 10^{9}$\Msolar, providing \co~measurements for 532 galaxies in combination with the original COLD GASS sample. With this large set of galaxies, \citet{sai17} found that the offset from the star forming main sequence is correlated strongly with both the molecular gas to stellar mass ratio (defined as \fh2 = \Mh2/\sm, which we simply refer to as molecular gas fraction for the rest of this paper) and the star formation efficiency (SFE = SFR/\Mh2). In other words, the elevation or suppression of star formation rate not only depends on the availability of gas but also on the conditions in the interstellar medium \citep[see also][]{hua14,tac13,tac18,pio20}. The ALLSMOG (APEX low-redshift legacy survey for molecular gas) survey \citep{bot14,cic17} complements the COLD GASS sample in the low mass regime by observing a further 88 nearby star-forming galaxies with the stellar masses between $10^{8.5}$ and $10^{10}$\Msolar~ using the APEX telescope. The combination of xCOLD GASS and ALLSMOG samples reveals that the CO luminosity not only strongly correlates with stellar mass and SFR but also varies with other factors, such as metallicity and \hi~ mass \citep[hereafter \MHI;][]{cic17}.

While surveys such as xCOLD GASS and ALLSMOG enable the exploration of the connection between the global gas content and SFR across a wide range of galaxy properties, as well as the variation of gas properties with respect to their positions on the SFR--\sm~ plane, details of the physical processes that shape the SFMS and its evolution  remain unclear. To take a step further, spatially resolved observations are required for three main reasons: 1) There is growing evidence that the SFR also traces \sm~ at kpc scales \citep{san13,wuy13,can16,hsi17,abd17,pan18,ell18,med18,vul19,can19,wan19,mor20}. This relation has been dubbed the `resolved' SFMS (rSFMS) and its existence suggests that the well-known global star forming main sequence may be an ensemble  effect of local processes; 2) Star formation takes places within giant molecular clouds (GMCs), whose sizes are on the order of hundreds of pcs or even smaller. The star formation law that characterizes the relation between SFR and gas density, often referred to as the Schmidt-Kennicutt relation \citep[or SK relation,][]{sch59,ken98}, is also found to vary with the substructures of galaxies and galactocentric radius \citep[e.g.,][]{big08,ler13,use15,rah16,uto17,sch19,che20}; 3) The spatial sequence of quenching (inside-out vs. outside-in vs. global) obtained from Integral Field Spectroscopy (IFS) observations provides powerful constraints on quenching mechanisms \citep[][also see S{\'a}nchez 2020 for a review]{gon14,gon16,tac15,li15,ell18,san18,lin19a}, particularly when combined with resolved gas observations \citep{lin17}. Obtaining a full picture of how the star formation is related to the global properties of galaxies and how it is quenched therefore requires mapping stars and gas in galaxies with sufficient spatial resolution and sensitivity. 

Recent spatially-resolved observational programs, such as the EDGE (the Extragalactic Database for Galaxy Evolution)-CALIFA (Calar Alto Legacy Integral Field Area) survey and the PHANGS (Physics at High Angular resolution in Nearby GalaxieS)-MUSE survey, are designed to combine the power of optical IFS and millimeter-wave interferometry observations to address key questions in star formation. The EDGE-CALIFA survey \citep{bol17} observed 126 CALIFA-selected galaxies in \co~and \coprime~using the Combined Array for Millimeter-wave Astronomy (CARMA) with an spatial resolution ($\sim$ 1.4 kpc) matched to CALIFA. This dataset enables studies of the relationships between molecular gas, stellar mass, star formation rate, metallicity, and dust extinction on kpc scales and their dependence on the global galaxy properties in a fairly large and representative local sample \citep[e.g.,][]{uto17,col18,dey19,bar20}. The PHANGS-MUSE project observed 19 nearby galaxies selected from a subset of the PHANGS-ALMA survey (A. K. Leroy et al. 2020, in preparation), utilizing the Multi Unit Spectroscopic Explorer (MUSE) at the VLT. Both ALMA and MUSE observations achieve an angular resolution $\sim$ 1 \arcsec, which corresponds to a physical scale of $<100$ pc in their sample. Although modest in sample size, PHANGS-MUSE therefore offers exquisite spatial resolution.  Combined, EDGE-CALIFA and PHANGS-MUSE offer complementary strengths for studying gas and star formation in the nearby universe.

While both EDGE-CALIFA and PHANGS-MUSE offer exceptional opportunities to study physical processes that govern star formation at (sub)kpc scales, these samples mostly lie on the star-forming main sequence and hence do not cover a sufficiently wide range in specific star formation rates (sSFR) to study the full gamut of processes from starbursts to quenching. In order to systematically understand the processes driving and regulating star formation, it is desirable to also include galaxies that are beyond the main sequence (in both directions). To test the feasibility of detecting resolved (kpc-scale) CO in green valley galaxies, in \citet{lin17} we conducted a pilot ALMA study of three MaNGA-selected galaxies. It was found that the molecular gas fraction \footnote{In \citet{lin17}, the molecular gas fraction is defined as \Mh2/(\sm+\Mh2), slightly different from \Mh2/\sm~used in this paper. The difference, however, is small as \Mh2~is in general less than 10\% of \sm~in our sample.} and SFE respond differently between bulge and disk regions as galaxies move away from the main sequence: the molecular gas fraction shows a stronger decline with respect to sSFR in bulges than in disks whereas SFE is reduced in both the bulge and disk regions \citep{lin17}, consistent with inside-out quenching. 

In this paper, we introduce the ALMA-MaNGA QUEnching and STar formation survey (ALMaQUEST), which expands the pilot sample of \citet{lin17} by more than an order of magnitude, covering not only main -sequence galaxies, but also starburst and green valley galaxies, all selected from the MaNGA survey. ALMaQUEST was designed to provide an extensive picture of the relationship between stellar populations, SFR, and gas at kpc scales by taking advantage of IFS data from MaNGA and resolved \co~observations from ALMA. Some of the main scientific questions that we aim to address with the ALMaQUEST survey include: 1) How are the properties of a galaxy's gas content linked to the resolved SFMS? 2) What are the primary physical mechanisms responsible for quenching? 3) Are starbursting galaxies driven by elevated \fh2~or enhanced SFE? 

This work is a presentation of the ALMaQUEST data that can be used as a companion to the various science papers. In \S 2, we describe the survey design, sample selections, and the data products used in this work. \S 3 characterizes both the global and local molecular gas properties of the sample. Some example science cases enabled by this sample are described in \S 4. A summary of this work is given in \S 5.

Throughout this paper we adopt the following cosmology: \textit{H}$_0$ = 70~\kms Mpc$^{-1}$, $\Omega_{\rm m} = 0.3$ and $\Omega_{\Lambda } = 0.7$. We use a Salpeter initial mass function (IMF). 
%All magnitudes are given in the AB system.

\section{SAMPLE and OBSERVATIONS \label{sec:data}}

\subsection{Sample Selection}

The ALMaQUEST survey consists of ALMA \co~ observations of galaxies selected from the MaNGA survey \citep{bun15,yan16b}. ALMaQUEST compiles datasets from four individual ALMA programs--2015.1.01225.S, 2017.1.01093.S, 2018.1.00558.S (PI: Lin), and 2018.1.00541.S (PI: Ellison). It contains 46 unique galaxies with a wide range of sSFR, spanning from the the green valley (hereafter GV), main sequence (hereafter MS), and up to the starburst (hereafter SB) regimes. While the majority of the targets are selected according to their global sSFR, the 12 SB galaxies are required to lie on or above the main sequence and show elevated SFR relative to the control sample within 0.5 \Re~ by at least 50 percent \citep[see][for details]{ell20a}. All of these observations adopt identical observing setups and reduction procedures. %Among all the 47 targets in ALMaQUEST, only one galaxy (plate-IFU 8616-1902) is not detected in CO. 
The locations of the 46 galaxies and all the MaNGA Data Release 15 (DR15) galaxies in the global SFR and \sm~ plane are shown in Figure \ref{fig:global_sfrsm_p3d}. The global measurements of SFR and \sm~are taken from the PIPE3D \citep{san16a,san16b} value-added catalog \citep{san18}, which sums the MaNGA spaxel measurements across the data cubes. Some basic quantities of the 46 ALMaQUEST galaxies are given in Table \ref{tab:sample}. Key information and characteristics of the  ALMaQUEST survey can be found in the ALMaQUEST webpage: http://arc.phys.uvic.ca/$\sim$almaquest/.

\subsection{MaNGA Data}
MaNGA is an IFS survey conducted with the SDSS 2.5m telescope \citep{gun06}, as part of the SDSS-IV survey \citep{alb17,bla17}. The MaNGA parent sample is selected to form an approximately-uniform distribution in $i$-band absolute magnitude, which corresponds to a roughly flat distribution in $log$ $M_{*}$ \citep{wak17}. MaNGA uses the BOSS spectrographs \citep{sme13} and couples them with hexagonal fibre bundles of different sizes \citep{dro15}. Each spectrum covers a wavelength range  of 3500-10,000\AA~ with a spectral resolution $\sim$60 kms$^{-1}$. After dithering, MaNGA data have an effective angular resolution (full width at half maximum; FWHM) of 2.5\arcsec \citep{law15}, corresponding to a physical scale of 0.5-6.5 kpc. Data cubes are grided with 0.5\arcsec spaxels. The methods used for sky subtraction and spectrophotometric calibration are described in \citet{law16} and \citet{yan16a}, respectively.

The MaNGA data used in this work is based on the SDSS Data Release 15 (DR15) version processed by the MaNGA reduction pipeline \citep{law16}. Measurements of the \sigsm~ and emission-line fluxes are taken from the public PIPE3D data products \citep{san16a,san18}. The stellar mass is obtained based on the best-fit stellar population model that describes the stellar continuum of a given spectrum. The best-fit stellar continuum is then subtracted from the reduced spectrum in order to obtain the emission line measurements. All the emission lines were extinction corrected using the Balmer decrement computed at each spaxel and a Milky Way
extinction curve with Rv = 3.1 \citep{car89}. The SFR is estimated based on this extinction corrected \ha~ flux following the
conversion given by Kennicutt (1998) with a Salpeter IMF. \sigsm~ and \sigsfr~ are computed using the stellar mass and SFR derived for each spaxel, normalized to the physical area of one spaxel with an inclination correction derived using the axis ratio from the NASA Sloan Atlas (NSA) catalog \footnote{http://nsatlas.org/}.
%\footnote{The recipe for coverting the \ha~luminosity into SFR may vary depending on the science goals. The procedure described here specifically applies to the studies presented in this work.  Alternative treatments, if any, are given in individual ALMaQUEST papers.}
We classify each MaNGA spaxel into regions where the dominant ionizing source is star formation, LI(N)ER , or Seyfert using the BPT diagnostic based on the [OIII]/\hb~vs. [SII]/\ha~ line ratios \citep{kew01,kew06}. We require a signal to noise (S/N) > 3 for the \ha~and~\hb~ lines and S/N > 2 for the [OIII] and [SII] lines when performing this classification.

\subsection{ALMA Observations}\label{sec:alma}
%%%%% Here we describe the observational setup and what we did in the visibility domain.
Molecular gas observations in $^{12}$CO(1-0) (rest frame 115.271204 GHz) were carried out
with ALMA during Cycle 3, 5, and 6 using the Band 3 receiver. The  observations were taken in the C43-2 configuration (synthesized beam FWHM $\sim$ 2.5$\arcsec$), thus matching the MaNGA resolution.
We used a single pointing with a field-of-view (FOV) of $\sim$ 50$\arcsec$.
The largest structure that we expect to be sensitive to is about 23$\arcsec$ ($\sim$14 kpc).
Our spectral setup  includes one high-resolution spectral window ($\sim$ 10 \kms) targeting $^{12}$CO(1-0), and one to three low-resolution  spectral window(s) ($\sim$ 90 \kms) around the target line aimed at detecting/studying the continuum. On-target integration time varies from 0.2 to 2.5 hours and is set to have a S/N (CO) greater than 3 for more than 50\% of spaxels with S/N (\ha) > 3. This setup was empirically shown in our pilot program \citep{lin17} to be a good compromise between the required integration time and the number of spaxels sufficient for carrying out statistically meaningful analyses. The actual sensitivity achieved for individual sources is given in Table \ref{tab:sample}.  
The  data   were calibrated using the ALMA data reduction software CASA version 4.5 or 5.4 \citep[Common Astronomy Software Applications;][]{mcm07} and standard ALMA pipeline\footnote{7977-12705, 7977-3704, and 7977-9101 were calibrated with CASA version 4.5 because they were observed in an earlier cycle.}. The systematic flux uncertainty associated with the calibration is typically 5 -- 10\% in Band 3.
Continuum is subtracted from the data in the visibility domain. Out of 46 ALMaQUEST objects, 4 (PLATEIFUs: 8084-3702, 8155-6101, 8615-3703, 8655-3701) are detected in the continuum.

%%%%% Here we form images (i.e., data cubes) from visibilities. Beam size and sensitivity are measured in image domain, so I moved them to here.
The task CLEAN was employed to clean the continuum-subtracted data down to 1$\sigma$ and  produce spectral line data  cubes  using Briggs weighting  with  a robust  parameter  of  0.5. 
The resultant native effective beamsize for each target ranges from 1.6$\arcsec$ to 2.8$\arcsec$.
28 out of 46 galaxies have a native effective beamsize comparable to the point spread function (PSF) of MaNGA (2.5$\arcsec$ $\pm$ 10\%), 14 galaxies have beamsize $\leqslant$ 2.3$\arcsec$, and 4 have beamsizes  $\sim$ 2.8$\arcsec$.
In order to facilitate the comparison between ALMA and MaNGA, we re-imaged the data and adopted a user-specified  pixel size (0.5$\arcsec$) and restoring beamsize (2.5$\arcsec$ $\times$ 2.5$\arcsec$) to match the image grid and the spatial resolution of the MaNGA images.
The final cubes have channel widths of 11 km s$^{-1}$ and rms noise ($\sigma_\mathrm{rms}$) of $\sim$ 0.2 -- 2 mJy beam$^{-1}$. The difference in $\sigma_\mathrm{rms}$ between the data cube with the original beamsize and our user-specified beamsize is as small as  $<$ 5\% because the original beamsize and the user-specified beamsize are  not significantly different.

\begin{figure*}
\centering
\includegraphics[angle=0,width=0.9\textwidth]{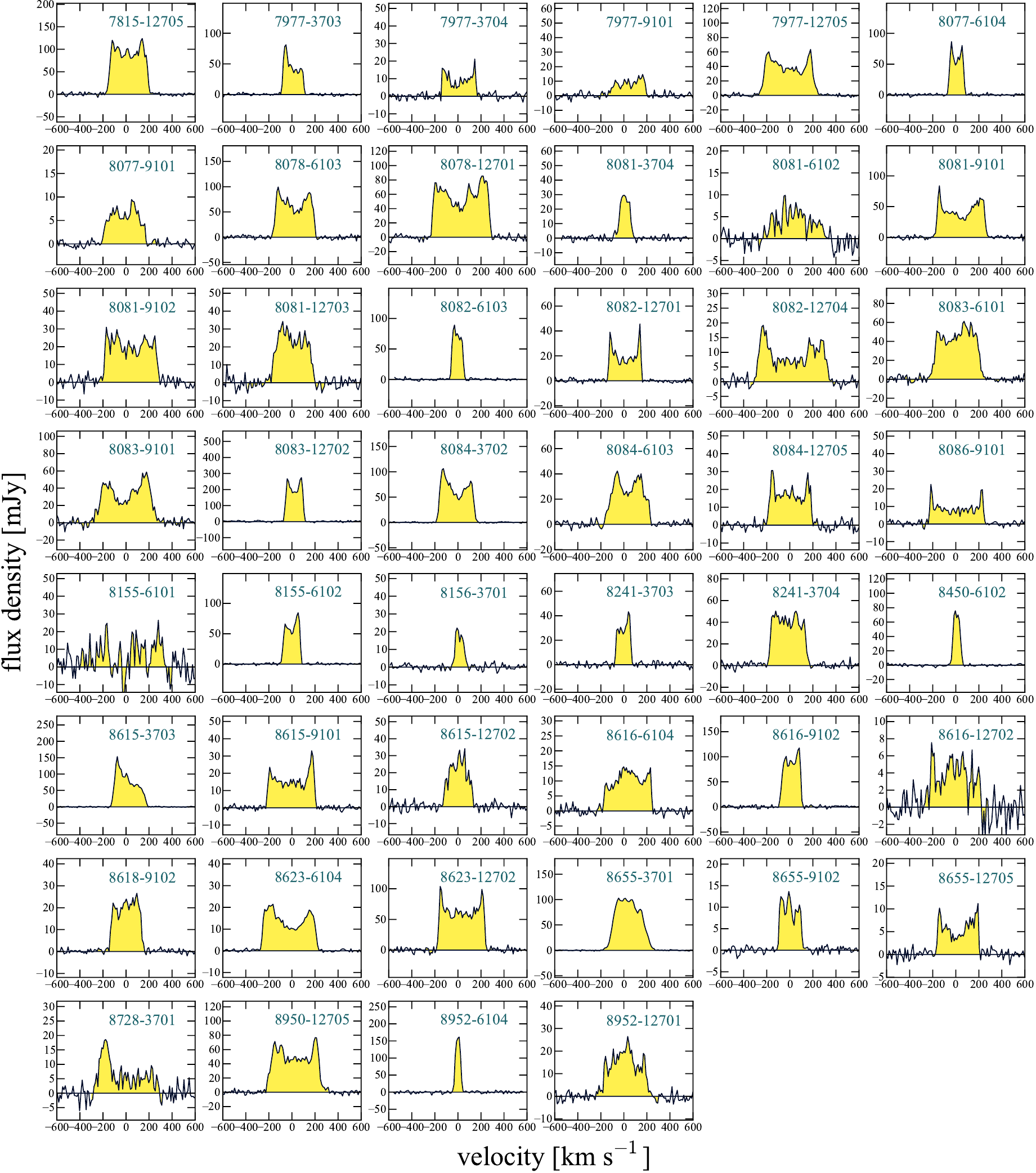}
\caption{\co~ spectra  integrated over the area enclosed by the 1.5 \Re~for 46 ALMaQUEST galaxies. The yellow shaded areas represent the region of the spectrum used for computing the line flux. The MaNGA plate-IFU identifier is given in the upper right of each panel. \label{fig:spectraA}}
\end{figure*}

%%%%% Here we present the spectra generated from the data cubes, and then the moment maps generated from the same data cubes.
In Figure \ref{fig:spectraA}, we show the 46 continuum-subtracted ALMA spectra centered on the position of the CO (1-0) line using the systematic velocity derived from the optical MaNGA redshift, integrated over the region enclosed by 1.5 effective radius (\Re). 
%The CO(1-0) line is clearly detected in all cases except for the galaxy 8155-6101 in which the detection is marginal. 
The 0th -- 2nd moment maps (0th: integrated intensity, 1st: intensity-weighted velocity field, and 2nd: intensity-weighted velocity dispersion) were constructed by using the task IMMOMENTS in CASA. 
The integrated intensity maps were created by integrating emission from a velocity range set by hand to match the observed line profile shown in Figure \ref{fig:spectraA} without any  clipping in signal.
Data in this velocity range were also used to generate velocity field and dispersion maps, with a 4$\sigma$ clipping applied to avoid noise contamination.

\subsection{GBT HI Observations}\label{sec:h1}
\hi-MaNGA is an \hi~ follow-up campaign for the MaNGA survey.
Complete details can be found in \citep{mas19} which we
briefly summarize here.

\hi-MaNGA uses the Green Bank Telescope to observe MaNGA galaxies at
$z<0.05$ lacking overlap with the ALFALFA survey \citep{hay18}.
The upper redshift limit is applied due to the declining sensitivity
beyond this redshift. Observations are conducted in position-switching
mode using the L-band receiver and VEGAS backend for a total (ON+OFF)
time of 30 minutes per target.  These integration times yield typical
rms noise levels of 1.5-2 mJy after boxcar and hanning smoothing to a
spectral resolution of $\sim$10 \kms.

\hi-MaNGA data are reduced (including RFI flagging, smoothing, and baseline
removal) using GBTIDL. The spectra are visually inspected for the
presence of an \hi~ emission line, and in the case of a detection the
flux, linewidth, and central velocity of the spectral line are
measured.  For non-detections, a 3$\sigma$ upper limit is estimated
assuming a linewidth of 200 \kms.  Fluxes and upper limits are
converted to \hi~ masses using  \MHI~ = 2.36e5$\times$(D / Mpc)$^{2}\times S_{21}$, where $S_{21}$ is the flux of 21cm emission line in units of Jy \kms.
Among the 46 ALMaQUEST galaxies, 26 and 7 galaxies are included in the \hi-MaNGA and ALFALFA samples, respectively. In this study, we use the \hi~ data from both the second data release of \hi-MaNGA
(D. Stark et al. in prep.) and the ALFALFA catalog \citep{hay18}.

\subsection{Global Measurements \label{sec:measurement}}
In Tables \ref{tab:property_15re} -- \ref{tab:property_manga}, we provide several key
measurements for the 46 ALMaQUEST galaxies, including the area enclosed within 1.5 \Re, stellar mass, star formation rate, CO flux, \h2~mass, sSFR, SFE, and \fh2. These integrated quantities are estimated in two ways: 1) By summing up the measured values over the area enclosed by 1.5 \Re~ and 2) by summing over the areas within the MaNGA bundles. These values are given in Table \ref{tab:property_15re} and Table \ref{tab:property_manga}, respectively. The choice of 1.5 \Re~is driven by the bundle coverage of MaNGA observations as two-third of the MaNGA sample is required to be covered by the MaNGA IFU out to 1.5 \Re~ and one-third of the sample is covered out to 2.5 \Re~\citep{wak17}.

The H$_{2}$ mass is computed from the CO flux by adopting a constant conversion factor ($\alpha_{\mathrm{CO}}$) of 4.35 \Msolar (K km s$^{-1}$ pc$^{2}$)$^{-1}$ \citep[e.g.,][]{bol13}. 
%Although it has been known that $\alpha_{\mathrm{CO}}$ may vary with the star formation rate and metallicity either globally or at sub-galactic scales \citep[e.g.,][]{bol13,mar13}, most of the calibrations are quantified in star-forming regions and the $\alpha_{\mathrm{CO}}$ in non-HII regions remain unclear. Therefore, we choose to report the results that are based on a constant value of the conversion factor throughout this work but 
We will discuss the effect of adopting a metallicity-dependent conversion factor in later sections (\S \ref{sec:global} and \S \ref{sec:local}). For the SFR measurement, only spaxels classified as star-forming using the [SII] BPT diagnostic \citep{kew01,kew06} are included in this study.  

%We have checked that the main trends presented in \S \ref{sec:global} remain similar if instead the total SFR is estimated using all spaxels or using an alternative BPT classification scheme based on the [NII] instead of [SII] lines \citep{kau03}.  

\begin{figure}
\centering
\includegraphics[angle=0,width=0.5\textwidth]{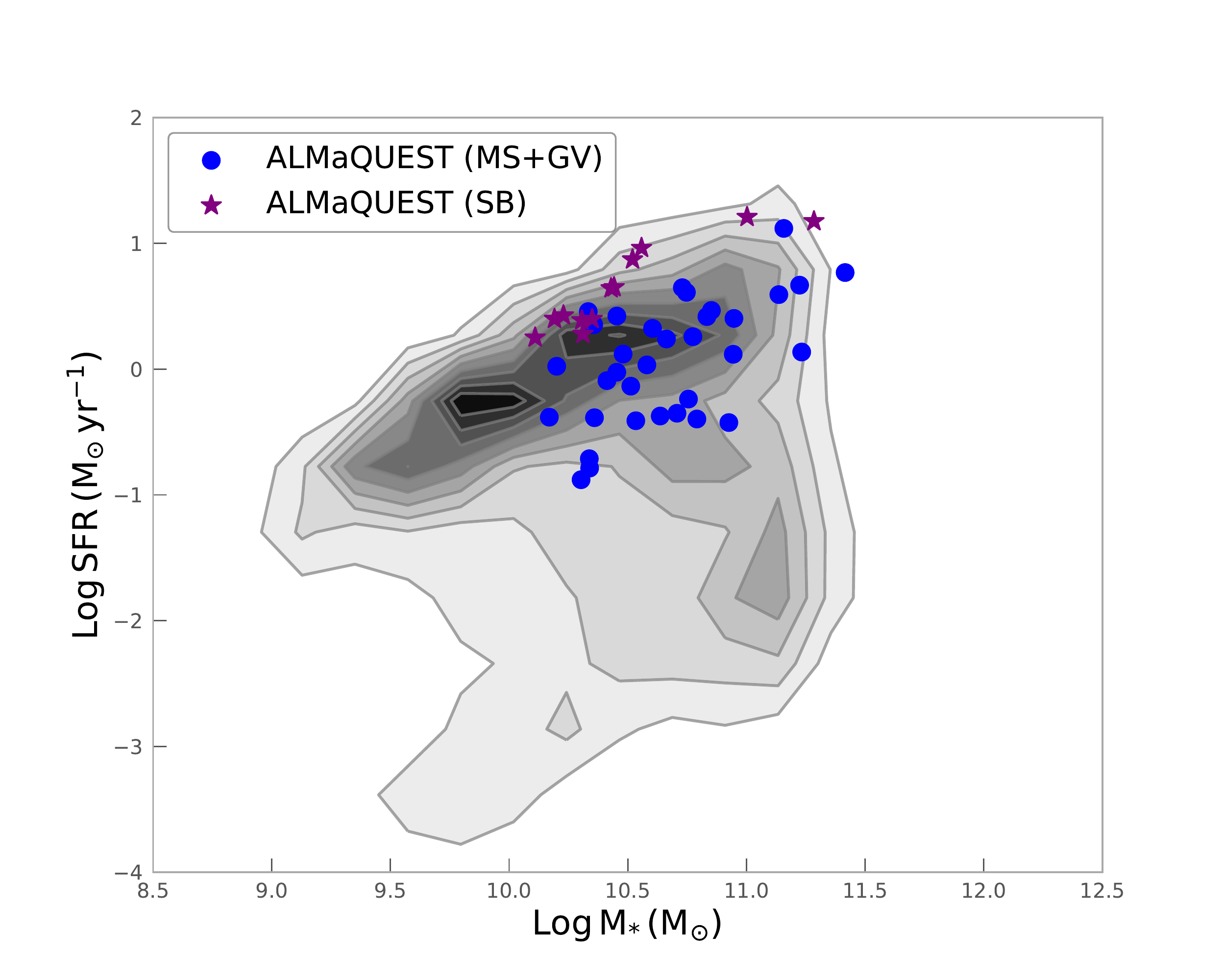}

%\plotthree{global_sfrsm.pdf}{global_sfrh2.pdf}{global_h2sm.pdf}
\caption{Distribution of the 46 ALMaQUEST galaxies (symbols) and all MaNGA DR15 galaxies (grey contours) in the global SFR and \sm~ plane. SFR and \sm~ are taken from the PIPE3D DR15 output. The purple stars and blue circles represent the starburst \citep{ell20a} and the remaining ALMaQUEST targets, respectively.  \label{fig:global_sfrsm_p3d}}
\end{figure}

\begin{figure}
\centering
\includegraphics[angle=0,width=0.45\textwidth]{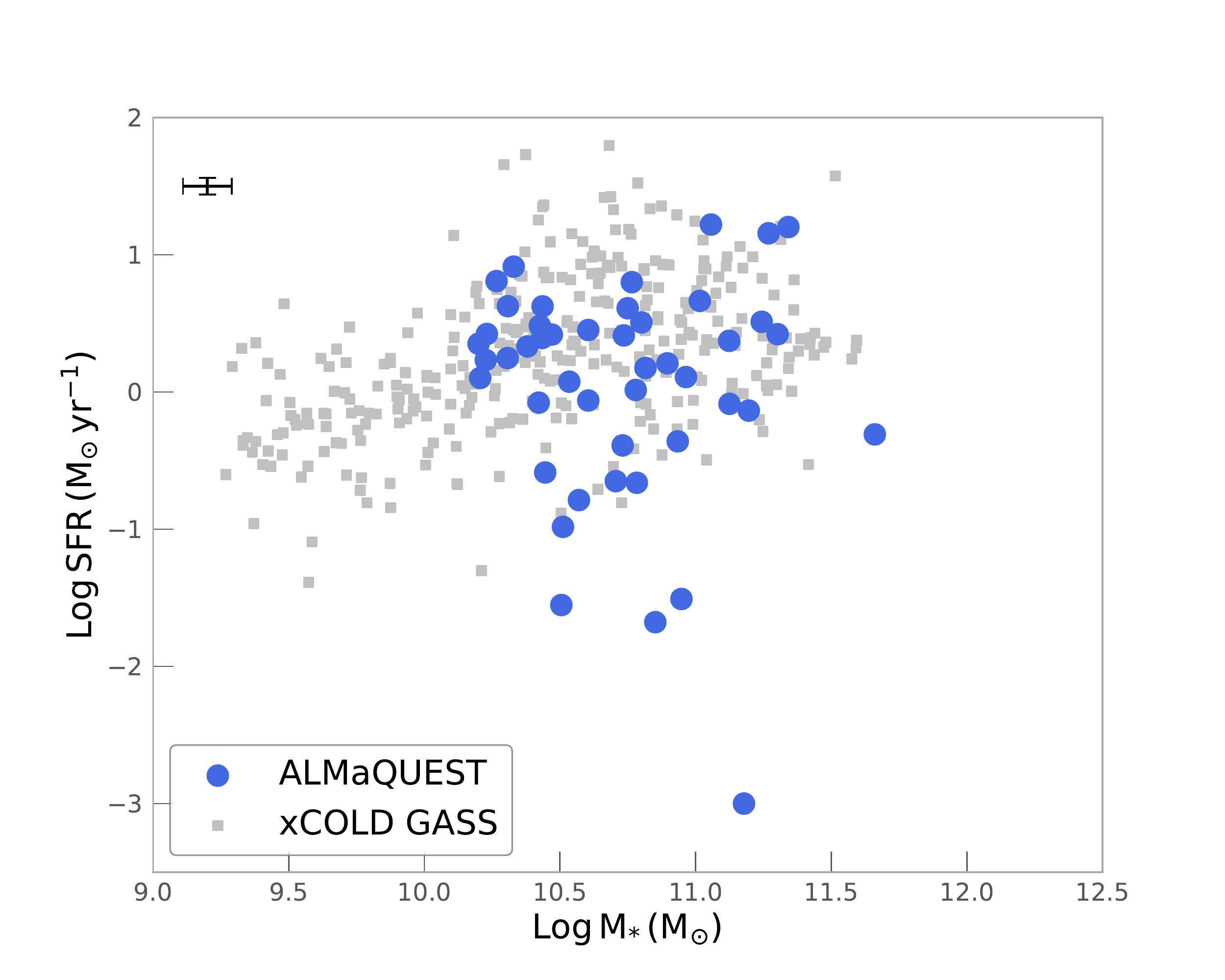}
\includegraphics[angle=0,width=0.5\textwidth]{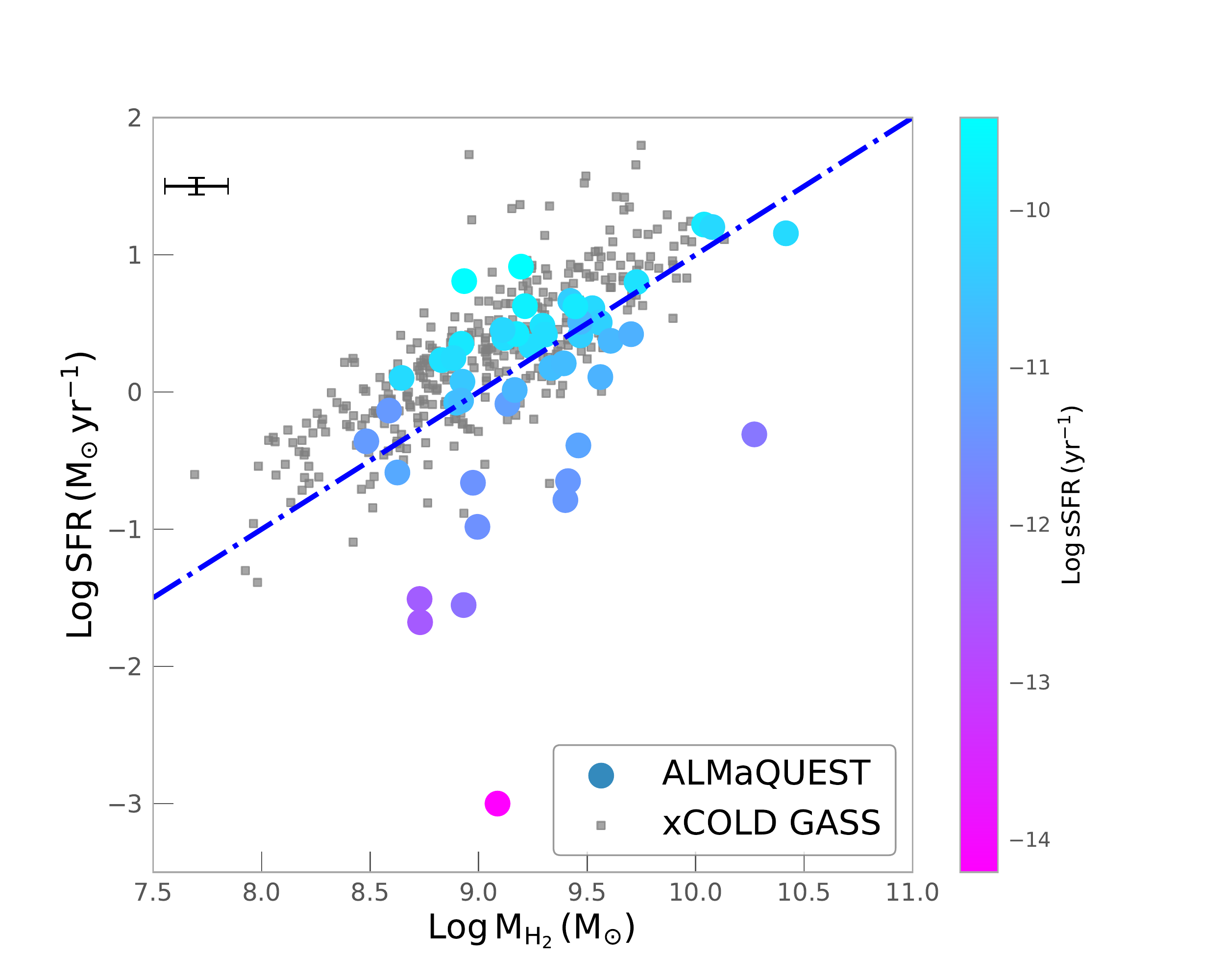}
\includegraphics[angle=0,width=0.5\textwidth]{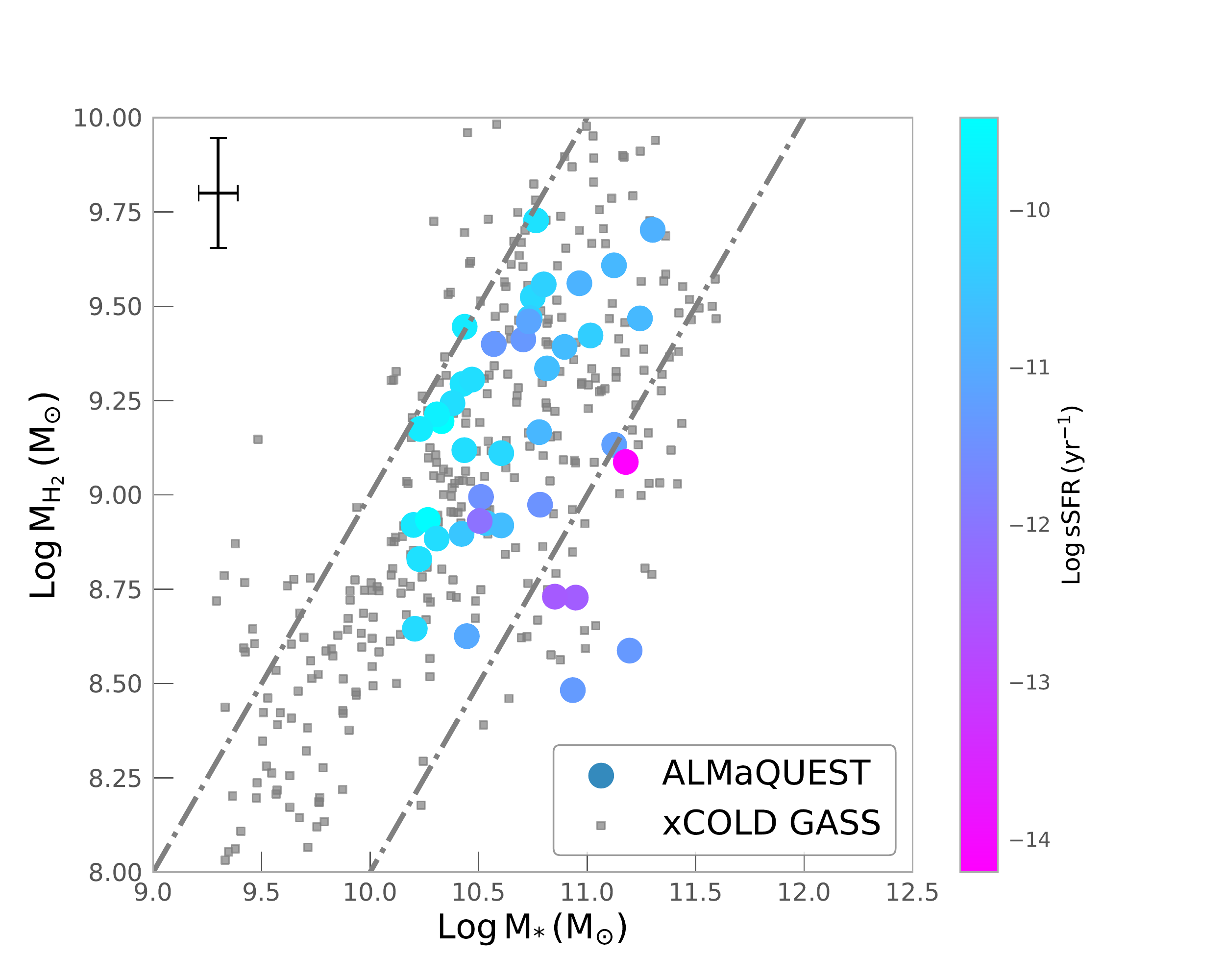}
%\plottwo{global_sfrh2_ssfr.pdf}{global_h2sm_ssfr.pdf}
\caption{Top panel: Distribution of the ALMaQUEST galaxies (blue symbols) in the global SFR and \sm~ plane. For comparison, the xCOLD GASS sample is shown as the grey squares and only galaxies with CO detections are shown for clarity. 
Middle panel: distribution of the ALMaQUEST galaxies in the global Schmidt-Kenicutt relation (SFR vs. \sm) color-coded by sSFR. The blue dot-dashed line corresponds to the SFE = 10$^{-9}$ yr$^{-1}$ line, not the best-fit to the data. Bottom panel: Relation between the global \Mh2~and \sm~(global molecular gas main sequence) of the ALMaQUEST sample color-coded by sSFR. The two grey dot-dashed lines, from left to right, show a gas-to-stellar ratio of 0.1 and 0.01, respectively. For the three panels, all of the ALMaQUEST quantities (i.e., SFR, \sm~, \Mh2) are computed within 1.5 \Re. The black error bars shown in the upper left corners represent the typical uncertainties.  \label{fig:global_2d_ssfr}}
\end{figure}

\begin{figure*}
\centering
\plottwo{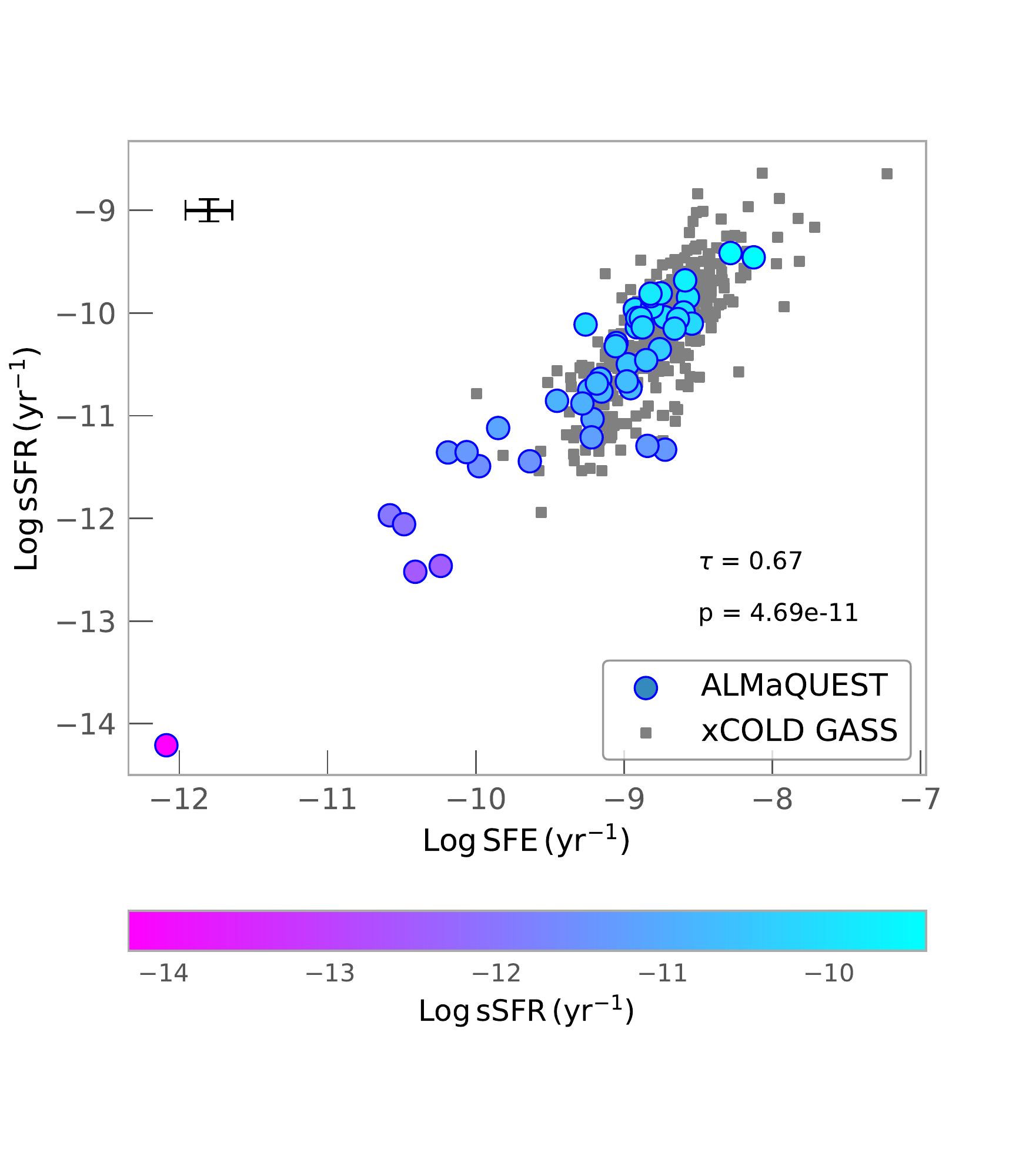}{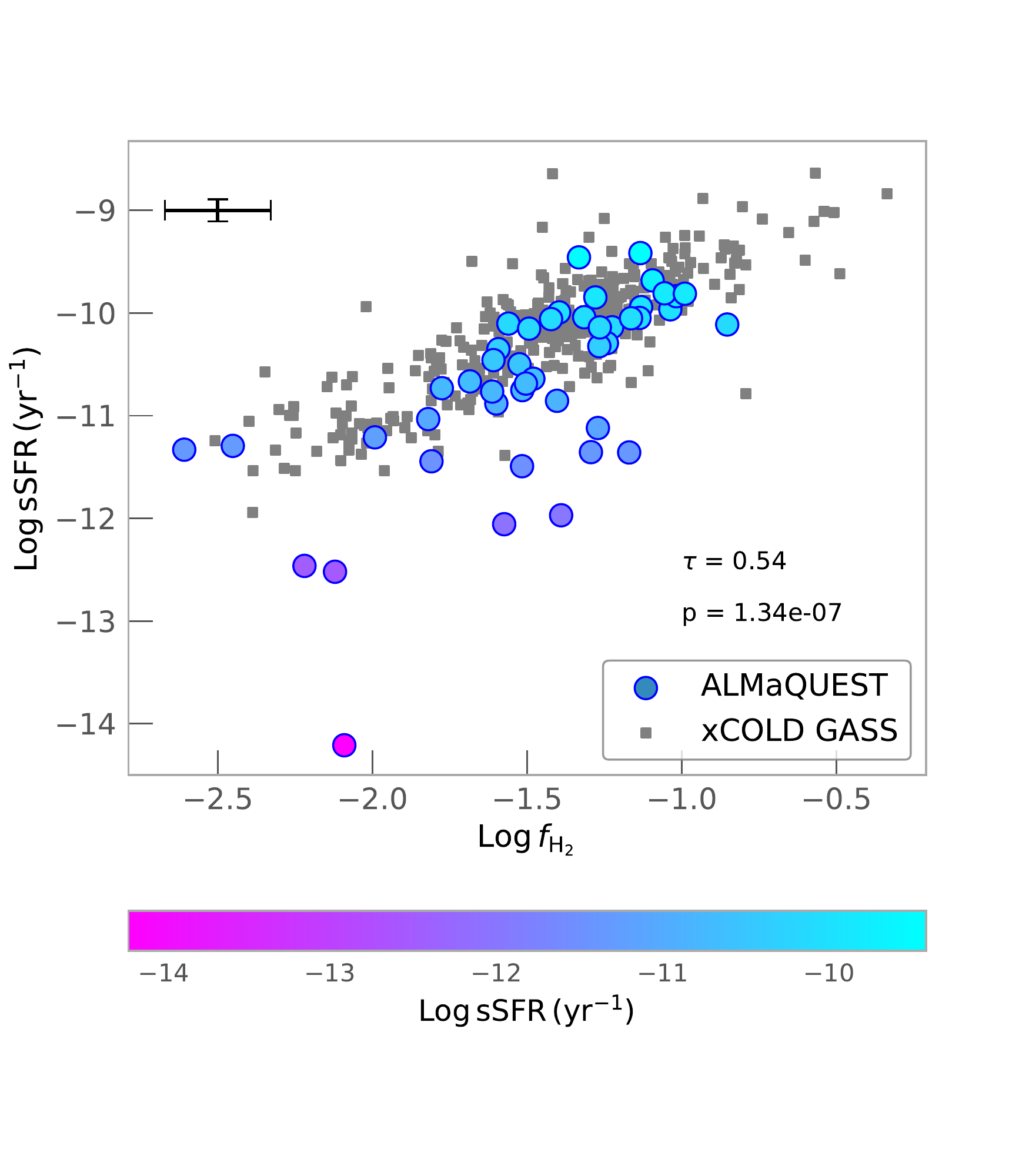}
\caption{The global sSFR of ALMaQUEST galaxies (blue circles) and xCOLD GASS (grey squares) as a function of SFE (left panel) and \fh2~(right panel). For both panels, all of the ALMaQUEST quantities (i.e., SFR, \sm~, \Mh2) are computed within 1.5 \Re. The black error bars shown in the upper left corners represent the typical uncertainties. The Kendall correlation coefficient ($\tau$) and $p$-value are shown in the legend.  \label{fig:global_sfefgas_ssfr}}
\end{figure*}

\begin{figure}
\centering
\includegraphics[angle=0,width=0.5\textwidth]{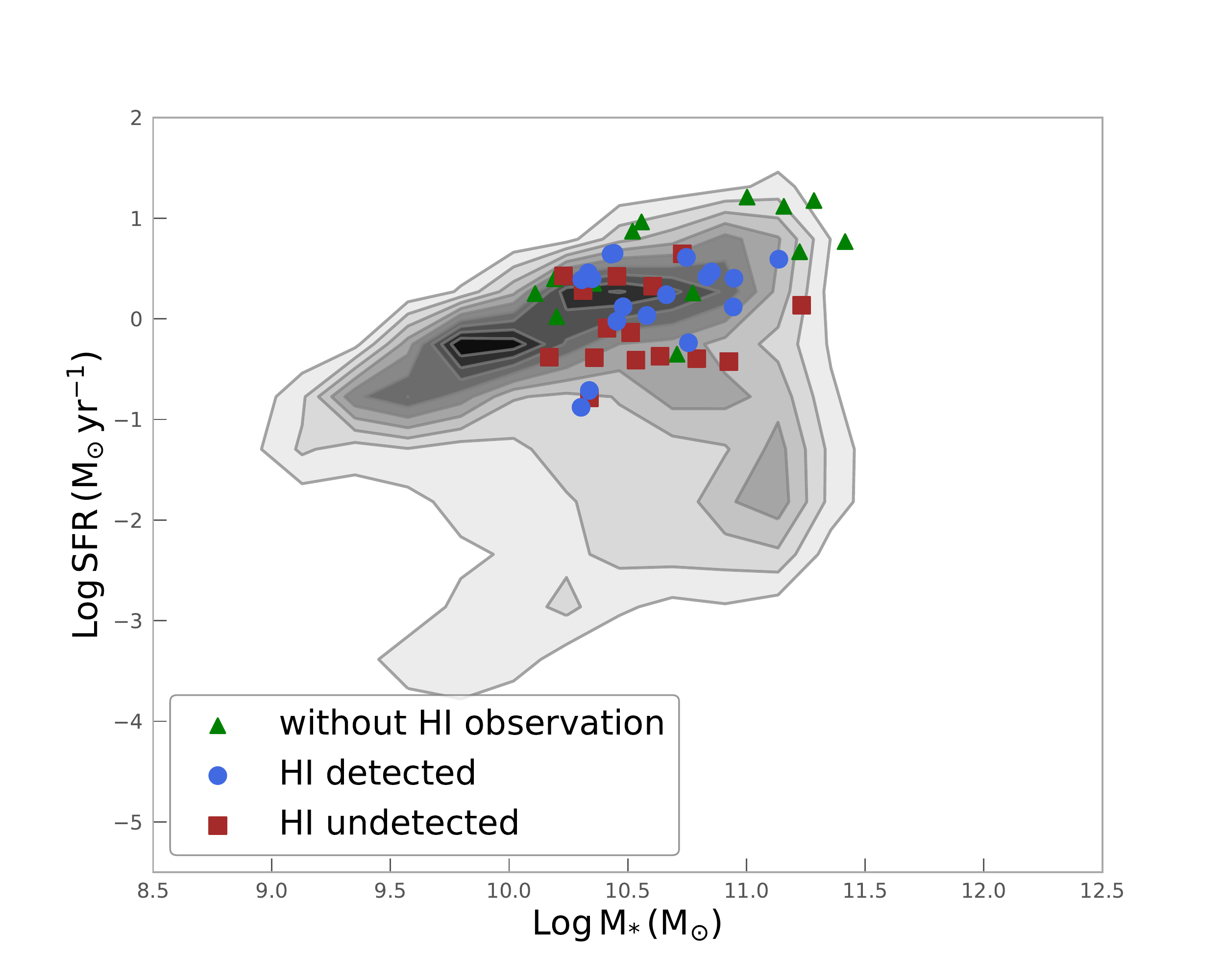}

%\plotthree{global_sfrsm.pdf}{global_sfrh2.pdf}{global_h2sm.pdf}
\caption{Distribution of the 46 ALMaQUEST galaxies (symbols) in the global SFR vs. \sm~ (taken from the PIPE3D DR15 output) plane, color-coded by their status of \hi~ observations (blue: with \hi~detection; brown: without \hi~ detection; green: no observation).  The full MaNGA DR15 galaxies are shown in grey contours.  \label{fig:global_sfrsm_p3d_h1}}
\end{figure}

\begin{figure*}
\centering
\includegraphics[angle=0,width=0.9\textwidth]{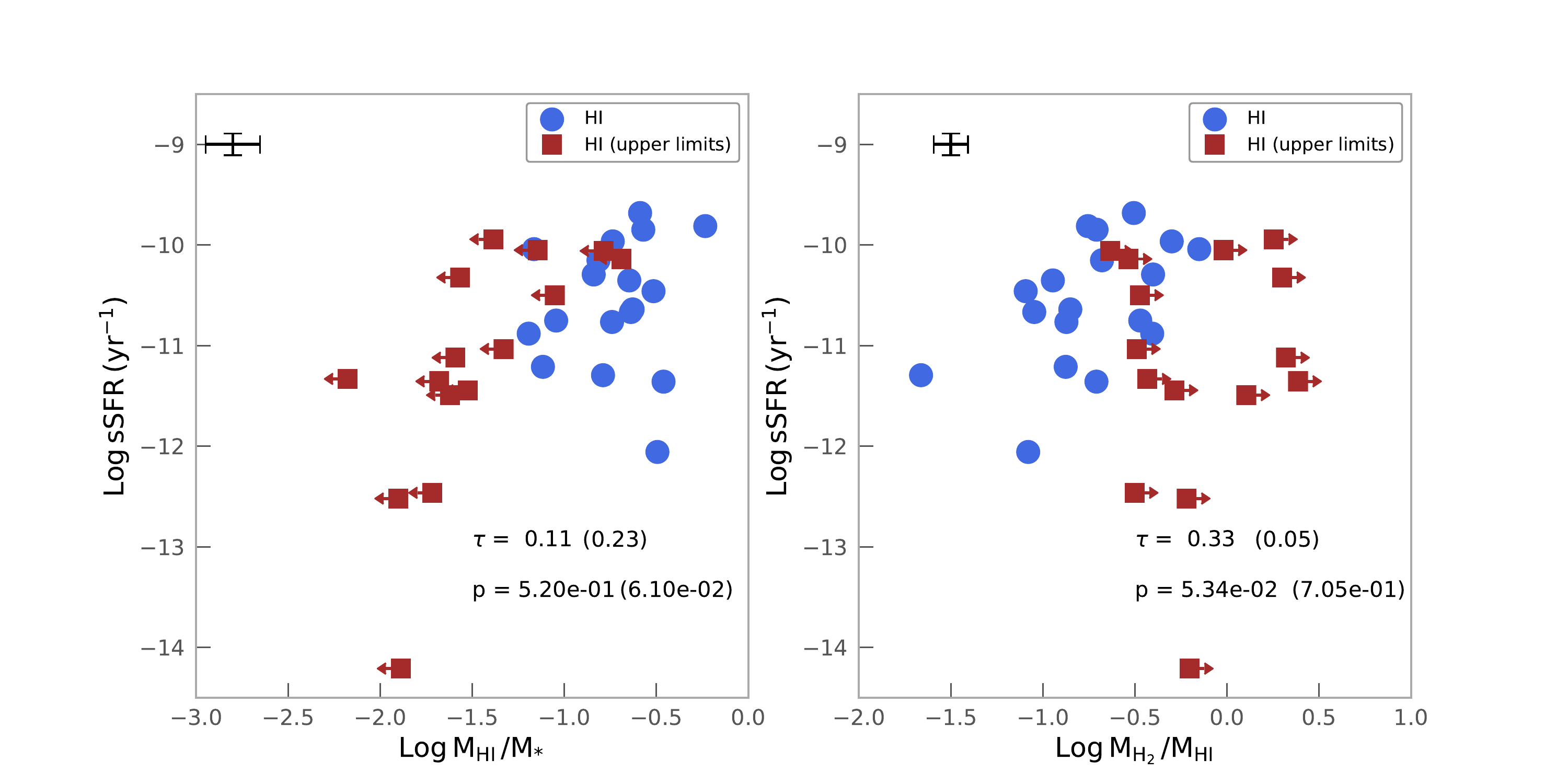}
%\plotthree{global_sfrsm.pdf}{global_sfrh2.pdf}{global_h2sm.pdf}
\caption{The global sSFR of ALMaQUEST galaxies as a function of the atomic gas fraction (left panel) and the \h2~to \hi~ ratio (right panel). The blue circles and brown squares represent datapoints with and without \hi~ detections, respectively. The black error bars shown in the upper left corners represent the typical uncertainties. The Kendall correlation coefficient ($\tau$) and $p$-value with/without the upper limits in \MHI~are shown in the legend (the values in the parenthesis are for the cases when considering the upper limits).  \label{fig:global_fh1_ssfr}}
\end{figure*}

%%\begin{figure*}
%%\centering
%%\includegraphics[angle=0,width=0.9\textwidth]{ALMaQUEST_moment_sncut3_all-1.pdf}
%%\caption{From left to right: SDSS $gri$ multicolor images, MaNGA raw \ha~intensity (10$^{-16}$ erg s$^{-1}$ cm$^{-2}$ per spaxel) , followed by ALMA \co~intensity  (Jy km s$^{-1}$ per beam), velocity (km s$^{-1}$), and velocity dispersion  (km s$^{-1}$) maps. The white circle in the lower-right corner of the CO panel illustrates the restoring beamsize. A S/N = 3 cut in the \co~intensity is applied when generating the associated velocity fields and dispersion maps.\label{fig:comap_1}}
%%\end{figure*}

%%%\figsetstart
%%%\figsetnum{7}
%%%\figsettitle{Images}

\figsetgrpstart
\figsetgrpnum{7.1}
\figsetgrptitle{ALMaQUEST targets}
\figsetplot{ALMaQUEST_moment_sncut3_all-1.pdf}
\figsetgrpnote{From left to right: SDSS $gri$ multicolor images, MaNGA raw \ha~intensity (10$^{-16}$ erg s$^{-1}$ cm$^{-2}$ per spaxel) , followed by ALMA \co~intensity  (Jy km s$^{-1}$ per beam), velocity (km s$^{-1}$), and velocity dispersion  (km s$^{-1}$) maps. The white circle in the lower-right corner of the CO panel illustrates the restoring beamsize. A S/N = 3 cut in the \co~intensity is applied when generating the associated velocity fields and dispersion maps.}
\figsetgrpend

\figsetend

\begin{figure*}
\figurenum{7}
\plotone{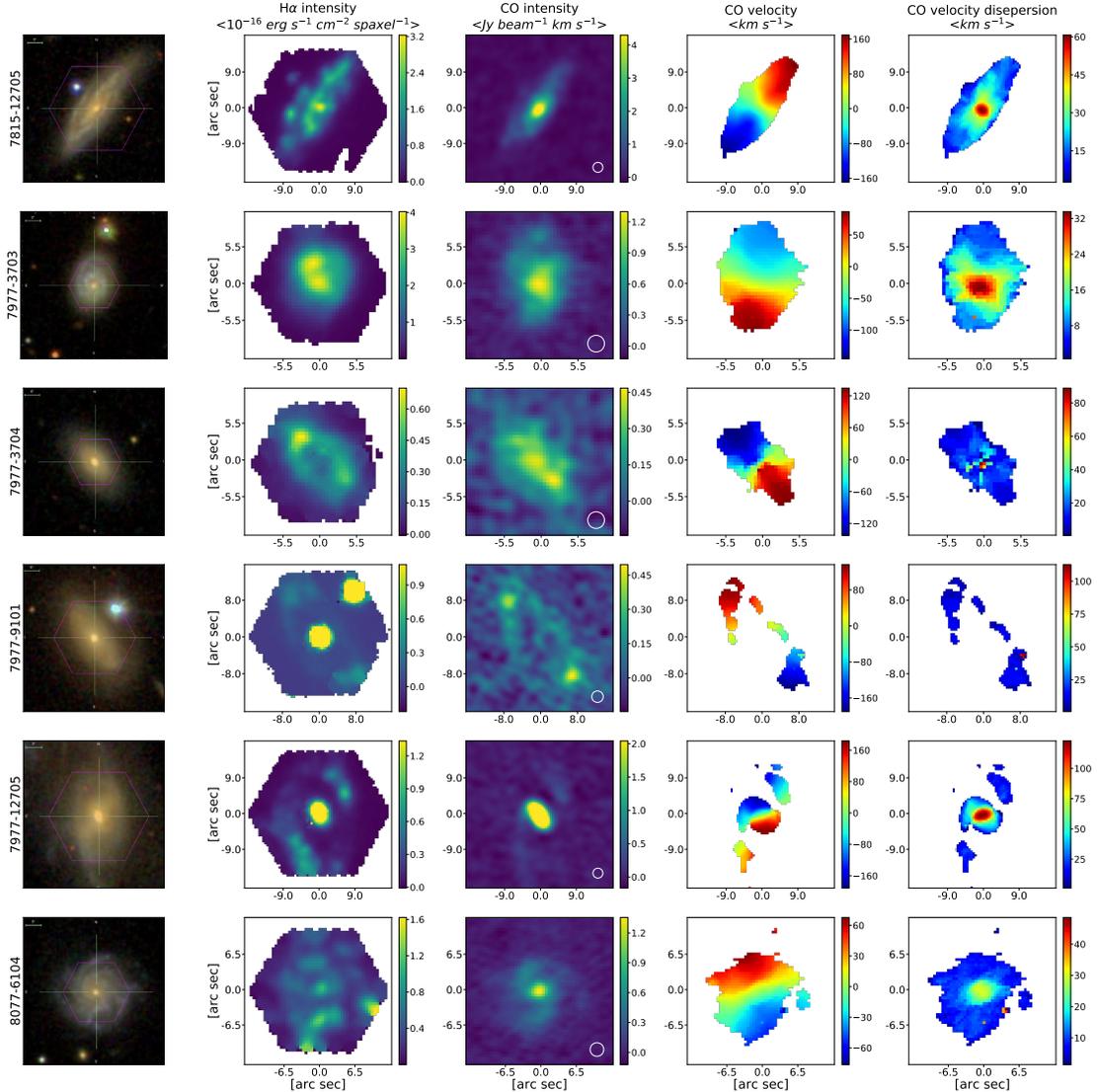}
\caption{Examples of ALMaQUEST targets. From left to right: SDSS $gri$ multicolor images, MaNGA raw \ha~intensity (10$^{-16}$ erg s$^{-1}$ cm$^{-2}$ per spaxel) , followed by ALMA \co~intensity  (Jy km s$^{-1}$ per beam), velocity (km s$^{-1}$), and velocity dispersion  (km s$^{-1}$) maps. The white circle in the lower-right corner of the CO panel illustrates the restoring beamsize. A S/N = 3 cut in the \co~intensity is applied when generating the associated velocity fields and dispersion maps. (The complete figure set (8 images) is available in the online journal.)\label{fig:comap_1}}
\end{figure*}

%%%\figsetstart
%%%\figsetnum{8}
%%%\figsettitle{Images}

\figsetgrpstart
\figsetgrpnum{8.1}
\figsetgrptitle{Physical products of the ALMaQUEST sample}
\figsetplot{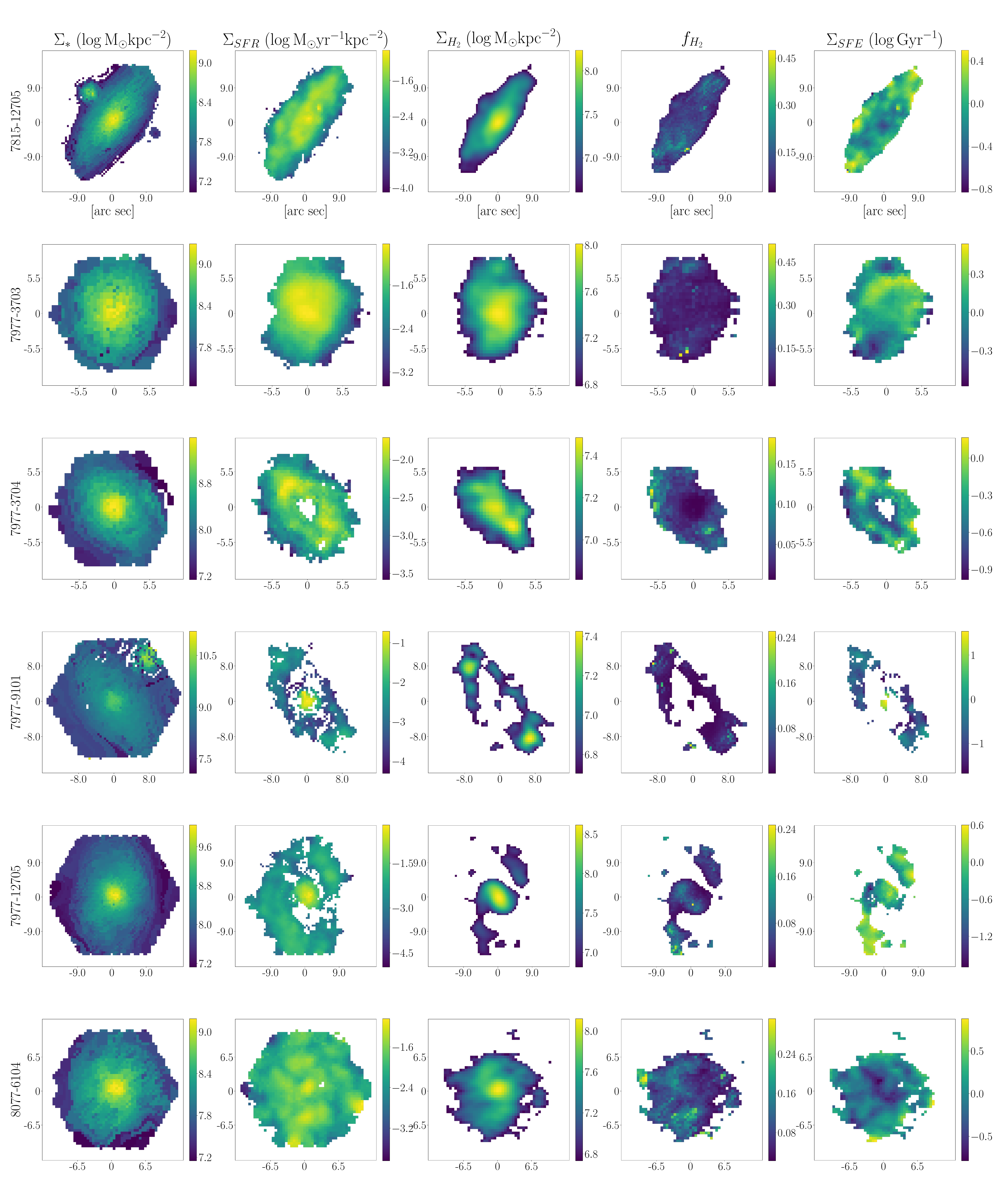}
\figsetgrpnote{From left to right: stellar mass surface density, \h2~mass surface density, SFR surface density, gas fraction, and star fromation efficiency. An S/N > 3 cut in the CO flux is applied. Only spaxels classified as the star-forming regions using the [SII] BPT criteria \citet{kew01,kew06} with an S/N > 3 cut in \ha~and \hb~lines and S/N > 2 cut in [OIII] and [SII] lines are shown in the SFR map.}
\figsetgrpend

\figsetend

\begin{figure*}
\figurenum{8}
\plotone{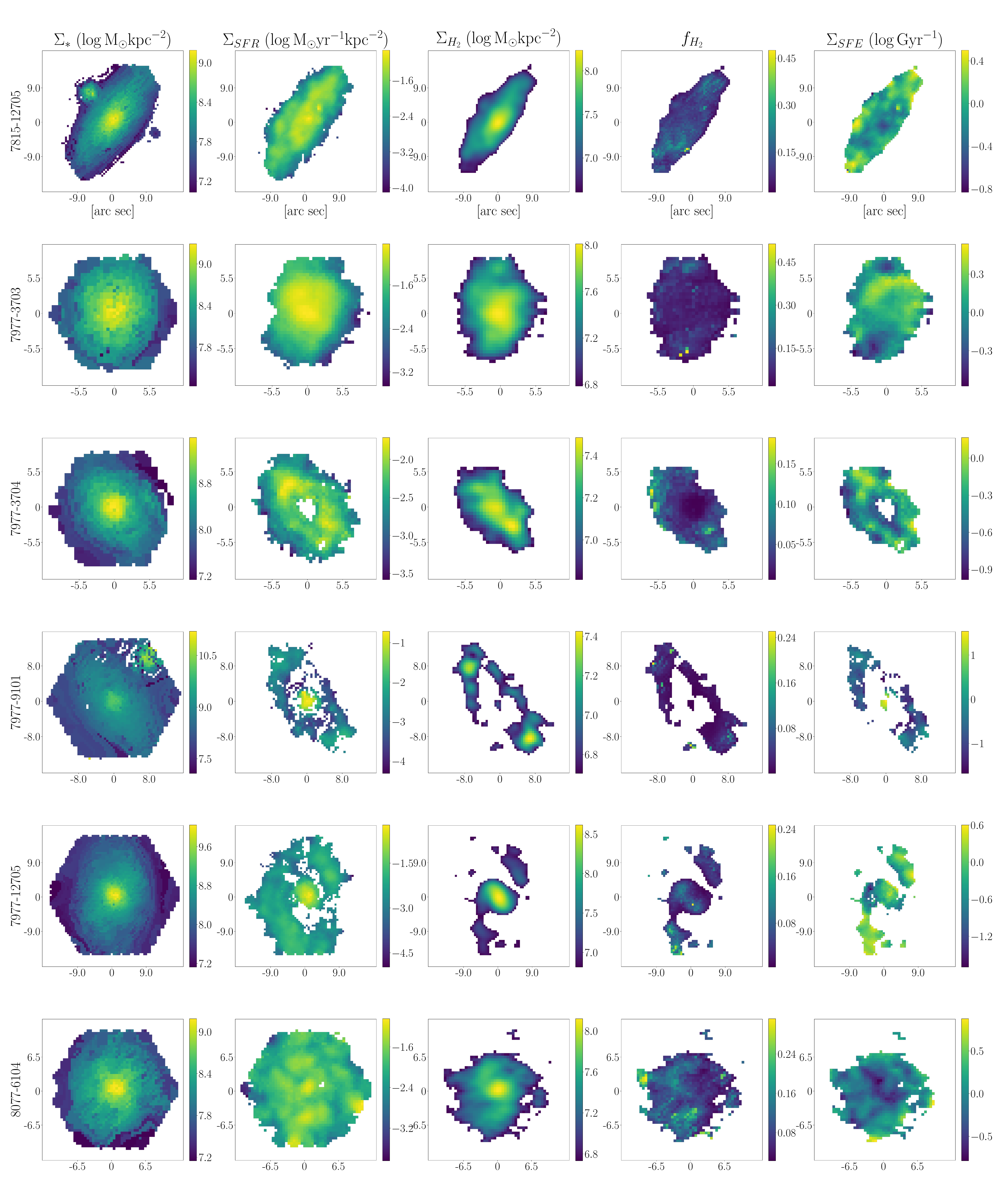}
\caption{Physical products of the ALMaQUEST sample. From left to right: stellar mass surface density, \h2~mass surface density, SFR surface density, gas fraction, and star fromation efficiency. An S/N > 3 cut in the CO flux is applied. Only spaxels classified as the star-forming regions using the [SII] BPT criteria \citet{kew01,kew06} with an S/N > 3 cut in \ha~and \hb~lines and S/N > 2 cut in [OIII] and [SII] lines are shown in the SFR map. (The complete figure set (8 images) is available in the online journal.)\label{fig:map_A}}
\end{figure*}

\section{RESULTS}
Having introduced the characteristics of the ALMaQUEST survey and the associated data products, in this section we present the molecular gas contents of ALMaQUEST galaxies and their relationships with the stellar populations, both globally and locally.

%The upper panel of Figure \ref{fig:global_2d} shows the `total' star formation rate vs. \h2~mass relation (upper panel) and the \h2~mass vs. stellar mass relation (lower panel) for various subsets of the ALMaQuEST sample. The blue line represents the gas depletion time (1/SFE) of 1 Gyr. It can be seen that starburst galaxies in general have SFE above the MS galaxies while GVs have SFE systematically below the MS. In contrast, the difference in the distribution of these three subsamples on the \Mh2~ and \sm~plane is less obvious. Although the averaged gas fraction decreases from SB to MS to GV, the range of the gas fraction among these subsamples is largely overlapped. Therefore, the global sSFR is mainly set by the variation in SFE rather than the gas fraction. 

\subsection{The integrated (global) scaling relations--SFR, \sm, and \Mh2 \label{sec:global}}

It has been pointed out that the position of galaxies in the global SFR--\sm~plane is driven by the combination of variations in both \fh2~ and SFE \citep[e.g.,][]{tac13,sar14,sai16,sai17}. In this subsection, we present the relations between the integrated quantities, i.e., the total SFR, total stellar mass, and the total \h2~mass, for the ALMaQUEST sample. Particularly, we investigate whether the variation in sSFR is primarily driven by the change in SFE or \fh2.
%In practice, there are several ways to compute the `total' quantities. 
%Because the spatial distributions and extensions of SFR, stellar mass, and CO differ within each galaxy, it is not straightforward to interpret the correlations by simply integrating the regions over the detection limit for each individual quantity. Since we are mainly interested in how these quantities are correlated with each other, an alternative is to integrate  the measurements over the same area. In light of this, we present the results that are measured within the areas enclosed by 1.5 \Re~ as given in Table \ref{tab:property_15re} (hereafter referred to as the aperture-matched measurements), although we note that the trends shown in Figure \ref{fig:global_2d_ssfr} remain unchanged if the MaNGA bundle-integrated properties (Table \ref{tab:property_manga}) are used. 
In Figure \ref{fig:global_2d_ssfr}, we present the correlations among the global SFR, \sm, and \Mh2, measured within the areas enclosed by 1.5\Re~ as given in Table \ref{tab:property_15re} (hereafter referred to as the aperture-matched measurements), although we note that the trends shown in Figure \ref{fig:global_2d_ssfr} remain unchanged if the MaNGA bundle-integrated properties (Table \ref{tab:property_manga}) are used. In all the three panels, the typical uncertainties in each measurement are shown in the upper-left corners. For SFR and \sm, the uncertainties represent the typical errors in the SFR and \sm~measurements, whereas the uncertainty in \Mh2~takes into account both the measurement error and the uncertainty in the CO-to-\Mh2~conversion factor with the latter being the dominant factor (see the end of this section). 

The top panel of Figure \ref{fig:global_2d_ssfr} shows the aperture-matched SFR versus \sm~(within 1.5\Re) for the ALMaQUEST sample (blue symbols). We also overplot the distributions of the xCOLD GASS sample \citep{sai17} shown as the grey symbols \footnote{We note that the xCOLD GASS measurements are not made strictly within 1.5Re and thus the data are not entirely comparable.}. An offset of 0.26 dex has been added to the xCOLD GASS SFR and \sm~measurements to account for the conversion from the Chabrier to the Salpeter IMF. For xCOLD GASS, only galaxies detected in CO are shown here for clarity.
The top panel of Figure \ref{fig:global_2d_ssfr} is analogous to the SFR versus \sm~ plot shown in Figure \ref{fig:global_sfrsm_p3d} but differs in the sense that the measurements in the latter are output by PIPE3D, which integrates values over the entire MaNGA bundle data cubes and uses all spaxels regardless their BPT classifications, whereas here we only consider star-forming spaxels within 1.5\Re. As a result, the SFR distribution is generally lower in Figure \ref{fig:global_2d_ssfr} compared to that in Figure \ref{fig:global_sfrsm_p3d}. Some galaxies with very few star-forming spaxels even have substantially lower sSFR compared to the PIPE3D output. 
It can be seen that while most of the ALMaQUEST galaxies remain in the MS and GV regimes, there are, however, a few ($\sim$5) galaxies falling into the quiescent population.

The middle panel of Figure \ref{fig:global_2d_ssfr} displays the global Schmidt-Kennicutt relation for the ALMaQUEST sample, color-coded by the sSFR. 
%We note that all spaxels within the desired area are used without removing non-HII regions classified by the BPT diagnostic. This is because This may lead to overestimation of the star formation rate, particularly for more evolved galaxies. However, as the ionizing ability of non-HII regions is in general one order of magnitude lower than the star formation \citep[e.g.,][]{hsi17} and the fraction of non-HII spaxels for most ALMaQUEST galaxies is below xxx\%, the effect of including non-HII regions on the SFR estimation is estimated to be less than xxx\%.  
The blue line represents a gas depletion time (1/SFE) of 1 Gyr to guide the eye. It can be seen that galaxies with higher (lower) sSFR tend to populate on the upper (lower) end of the SFR -- \Mh2 relation, suggesting a strong role of SFE in determining the sSFR of galaxies. This is in good qualitative agreement with previous studies on the global molecular gas content, which have found a strong relationship between SFE and sSFR \citep[e.g.,][]{hua14,sai17,bol17}.

In the bottom panel of Figure \ref{fig:global_2d_ssfr}, we show \Mh2~as a function of \sm, which is the global version of the molecular gas main sequence \citep[MGMS,][]{lin19b}. It was previously found that these two quantities for typical star-forming galaxies are almost linearly correlated with each other \citep[e.g.][]{cic17}. As our sample consists of both galaxies on and below the SFMS, we are able to explore how galaxies with different sSFR populate in this diagram. It can be seen that at a fixed sSFR, there appears to be a large spread in the molecular gas-to-stellar mass ratio, i.e. \fh2, and vice versa. The variation in sSFR with respect to \fh2~is hence less obvious compared to that with respect to SFE. We also notice that while the galaxies with sSFR < 10$^{-11}$   yr$^{-1}$ span a wide range in terms of the molecular gas-to-stellar mass ratio, their values extend to the regime below 1\%, suggesting a strong depletion of molecular gas within 1.5\Re~ in some of the galaxies with low sSFR. 
%In summary, the SB and MS samples are more distinct in their SFE rather than \fh2, similar to the trends seen in the resolved properties reported by \citet{ell20b}. On the other hand, both lower SFE and lower \fh2~are seen in the GV galaxies as opposed to SB and MS populations.

%Unlike MS and green valley galaxies, which are identified based on their global sSFR, the SB sample is selected according to their central elevated SFR and hence is not necessarily the ones with the highest global sSFR. To see how galaxies with different sSFR populate in the SFR versus \Mh2~ and \Mh2 versus \sm~planes, we again show these two relations but color-coded by the integrated sSFR in Figure \ref{fig:global_2d_ssfr}. 
To better quantify the relative contributions between SFE and \fh2~to the sSFR, Figure \ref{fig:global_sfefgas_ssfr} shows the correlations between sSFR with respect to SFE (left panel) and \fh2~(right panel). The ALMaQUEST and xCOLD GASS measurements are shown in blue and grey symbols, respectively. It can be seen that both samples follow very similar trends, although the ALMaQUEST survey shows several outliers from the sSFR vs. \fh2~relation. This is likely due to the fact that the CO-detected sample of the ALMaQUEST extends to lower sSFR regime.
 The Kendall correlation analysis for the ALMaQUEST sample shows that the correlation between sSFR and SFE ($\tau$ = 0.67) is slightly stronger compared to that between sSFR and \fh2 ($\tau$ = 0.54). However, one potential caveat in interpreting this result is that the SFE spans a much wider range (4 orders of magnitude) than the \fh2~does, owing to the inclusion of a datapoint with SFE $\sim$ -12.  To test if the tighter correlation seen in the sSFR vs. SFE is driven by the dynamical range effect, we repeat the analysis by excluding the lowest SFE datapoint. We find that the $\tau$ value only slightly decreases from 0.67 to 0.66 in this case and remains greater than the correlation between sSFR and \fh2. In summary, both SFE and \fh2~contribute to the variation in sSFR of galaxies, although SFE is found to play a slightly stronger role in governing the sSFR of galaxies. A similar trend has also been seen in previous studies, which found a comparable contribution of \fh2~ and SFE in the regulation \citep[e.g.][]{sai17,pio20}.

As described in \S \ref{sec:measurement}, the global SFR here are estimated using the star-forming spaxels classified with the [SII] BPT diagnostic \citep{kew01,kew06}. To test if our results are stable against different choices of area types included in the SFR calculation, in the Appendix we also show the results using all spaxels or an alternative BPT classification scheme based on the [NII] diagnostic \citep{kau03}. The former has a caveat that the total SFR can be overestimated since the non-star forming spaxels where the \ha~emissions powered by non-star formation mechanisms are also included. In both cases, we also see positive correlations between sSFR and SFE and between sSFR and \fh2. While the dependence of sSFR on SFE or \fh2~ are comparable in the case of all spaxels considered, the correlation with respect to SFE is found to be slightly stronger than with respect to \fh2~in the case where only [NII] BPT classified star-forming spaxels are included, in good agreement with the result based on the [SII] BPT method.

We note that our results described above are drawn based on a constant value of $\alpha_{\mathrm{CO}}$. To examine the
potential effect of this assumption, we also consider two types of varying $\alpha_{\mathrm{CO}}$, one taken from \citet{sun20}, which takes into account the metallicity dependence alone, and the other from \citet{nan12}, which considers the dependence on both the CO line intensity and metallicity. In the first method, we adopt Eq. (4) of \citet{sun20}:

\begin{equation}\label{eq: 2d}
\alpha_{\mathrm{CO}} = 4.35 (Z/Z_{\odot})^{-1.6}~ \rm M_{\odot} ~pc^{-2}(K~km~s^{-1})^{-1} ,
\end{equation}
where Z is the (linear) gas phase abundance and Z$_{\odot}$ is the solar value. The gas phase metallicity in log can be calculated through the O3N2 calibrator \citep{pet04}:
%The  be calculated through the O3N2-based gas phase metallicity \citep{pet04} at a given spaxel through:

%\begin{equation}\label{eq: 2d}
%\begin{aligned}
%\mathrm{log}~ \alpha_{\mathrm{CO}} = \mathrm{log} (4.35) - 1.6 \times ([12 + \mathrm{log (O/H)}]_{obs})\\ - 1.6 \times ([12 + \mathrm{log (O/H)}]_{solar})  ~(\mathrm{pc}^{-2}(\mathrm{K~km~s}^{-1})^{-1}),
%\end{aligned}
%\end{equation}
%where [12 + log (O/H)]$_{obs}$ and [12 + log (O/H)]$_{solar}$ are the observed gas-phase metallicity and the solar value (8.69), respectively. 

\begin{equation}
12 + \mathrm{log (O/H)} = 8.73 - 0.32 \times O3N2.
\end{equation}

In the second method, we utilize Eq. (11) of \citet{nan12}:
\begin{equation}
\alpha_{\mathrm{CO}} = \dfrac{\mathrm{min}[6.3, 10.7 \times  \mathrm{W}_{\mathrm{CO}}^{-0.32}]}{(Z/Z_{\odot})^{0.65}},
\end{equation}

where W$_{\mathrm{CO}}$ is the CO line intensity in units of K-\kms.
The metallicity is calculated for spaxels with S/N > 3 for \ha~and~\hb~ and S/N > 2 for \othree~and~ \ntwo. For spaxels that do not fulfill the above criteria, we set the metallicity to be the solar value. Although this approach may not be ideal, it offers an approximate view of the level of potential impacts on our derived results. 
%The $\alpha_{\mathrm{CO}}$ is computed spaxel by spaxel following the aforementioned two methods. 
For a given galaxy, we then compute $\alpha_{\mathrm{CO}}$ based on the median value of the spaxel-based metallicities. With the varying $\alpha_{\mathrm{CO}}$ applied, we find that over the entire ALMaQUEST sample, the difference in the global \h2~mass relative to the original global \h2~mass based on a constant value can be as large as 0.27 dex with an average value of $\sim 0.05$ and $\sim 0.14$ dex in method 1 and method 2, respectively. The latter is then combined with the measurement error to calculate the uncertainty of \Mh2 that is shown in Figure \ref{fig:global_2d_ssfr}. We then proceed with the same analysis and find that despite a systematic offset seen in the \Mh2~measurement, the varying $\alpha_{\mathrm{CO}}$ does not affect any trends in the global results presented in this section.

\subsection{The role of \hi~ gas \label{sec:global_HI}}

While the star formation is closely related to the molecular gas \citep{won02,gao04,wu05,del19}, the atomic gas, particularly the atomic hydrogen, dominates the cold gas mass budget on galactic scales and provides the fuel for future star formation. Although this paper focuses on the influence of molecular gas on star formation, a subset of the ALMaQUEST galaxies have been observed in \hi~ previously (see \S \ref{sec:h1}), which allows us to investigate the connection between gas and star formation from the atomic hydrogen point of view \citep[e.g.][]{bos14,cat18}. For example, one can study the dependence of sSFR on the atomic gas fraction (\MHI/\sm), which represents the availability of cold gas reservoir, as well as on the molecular-to-atomic gas ratio (\Mh2/\MHI), which is related to the efficiency of transforming the atomic gas into the molecular phase. Different from the CO observations, the \hi~ data were taken with single-dish telescopes and have a beam size that well exceeds the optical diameter of the galaxy. The \hi~ mass used here is hence the `total' \hi~ mass rather than the aperture-matched quantity, such as SFR, \sm, and \Mh2~used in the previous section.

There are 33 ALMaQUEST galaxies that overlap with the \hi-MaNGA \citep{mas19} and ALFALFA \citep{hay18} observations, of which 18 are detected in \hi~ with S/N > 5. Figure \ref{fig:global_sfrsm_p3d_h1} shows their distributions in the SFR vs. \sm~plane (blue: \hi~detected; brown: \hi~undetected) with respect to the rest 13 ALMaQUEST galaxies without \hi~observations (green triangles). Proportionally there are fewer high stellar mass galaxies with \hi~observations (blue + brown symbols) compared to the rest of the ALMaQUEST sample (green symbols). Among the \hi~sample, galaxies with and without \hi~detections are distributed similarly, albeit with small sample sizes.

We compute \MHI/\sm~and \Mh2/\MHI~ for all the 33 galaxies covered by the \hi~ observations and use the upper limits of \MHI~in the case of non-detection. According to the \Mh2/\MHI~ ratios, the 18 \hi-detected galaxies are all \hi-dominated (with respect to \Mh2) systems. In Figure \ref{fig:global_fh1_ssfr} we plot the dependence of sSFR on \MHI/\sm~and \Mh2/\MHI~ in the left and right panels, respectively.  We perform the Kendall correlation analysis for two cases, one only for the 18 galaxies with solid \hi~ detections and the other including the additional 15 galaxies with upper limits. In both cases (with/without upper limits), the dependence on \MHI/\sm~ and \Mh2/\MHI~ are weaker compared to the dependence of sSFR on SFE or \fh2~presented in Figure \ref{fig:global_sfefgas_ssfr}, suggesting that sSFR is more linked to the molecular gas budget than to the atomic gas. Nevertheless, it is worth noting that the dependence of sSFR on \MHI/\sm~ found in our sample is weaker than the results from earlier studies built upon larger samples \citep{sai16,cat18}. Furthermore, in our case without upper limits considered, the correlation is slightly stronger in the sSFR vs. \Mh2/\MHI~relation than in the sSFR vs \MHI/\sm~relation, whereas the trend is reversed when the upper limits are taken into account. This shows that galaxies without \hi~ detections have a significant impact on the strength of the correlation. Therefore, our results should be taken with caution given that the current ALMaQUEST \hi~ sample size is still limited. 

\subsection{Resolved properties of the ALMaQUEST sample \label{sec:local}}
As our main science goals of the ALMaQUEST survey rely on small scale properties of galaxies, we also constructed spatially resolved intensity and velocity maps for individual galaxies using the procedures described in \S \ref{sec:alma}.
In Figure \ref{fig:comap_1}, we show the SDSS optical image, MaNGA \ha~intensity, ALMA CO(1-0) integrated intensity, velocity, and velocity dispersion maps for the 46 ALMaQUEST galaxies. In general, it is evident from these maps that the \ha~flux does not always trace the CO(1-0) emission. Since \ha~ is a good tracer of SFR in regions where the photon ionization is dominated by star formation and CO(1-0) mass can be converted into the \h2~mass, the discrepancy in the spatial distributions between \ha~and CO(1-0) hence indicates a possible variation of SFE within a galaxy. 
%The non-uniform spatial distribution of SFE has also been shown in previous studies of nearby galaxies\citep[e.g.][]{hua15,uto17,col18,sch19}.

To further illustrate this point, we display maps for various physical quantities in Figure \ref{fig:map_A}, including stellar mass surface density \sigsm, \h2~mass surface density (\sigh2) using a constant $\alpha_{\mathrm{CO}}$ = 4.35, extinction-corrected star formation rate surface density (\sigsfr), \fh2, and SFE. Figure \ref{fig:map_A} shows that neither \fh2~nor SFE is a constant across a given galaxy, as also been shown in previous studies of nearby galaxies\citep[e.g.][]{ler08,hua15,uto17,col18,sch19,dey19}. To quantify the variation, we plot the histograms of sSFR, \fh2~ and SFE computed on a spaxel-by-spaxel basis for the star-forming spaxels of 46 ALMaQUEST galaxies in Figures \ref{fig:ssfr}, \ref{fig:fgas}, and \ref{fig:sfe}, respectively. In the bottom right panel of each figure (Figures \ref{fig:ssfr}--\ref{fig:sfe}), we show the combined distributions from all the 46 galaxies. When looking at the combined ensemble of spaxels, the distributions of all the three quantities (sSFR, \fh2~and SFE) are close to a Gaussian in log space. For the rest panels, the blue and green vertical lines show the median value of the histogram and the globally averaged value in a given galaxy, respectively. It can be seen that all of these quantities show a large spread (typically spanning 1-2 orders of magnitude within a galaxy), meaning that there is a strong variation even within a single galaxy. Some of the variations are found to be associated with the radial position of the galaxies while some are not (Pan et al. in prep.). Figures \ref{fig:ssfr}, \ref{fig:fgas}, and \ref{fig:sfe} therefore serve as a caveat that using a global value may not properly capture the `intrinsic' gas content of galaxies \citep[also see][]{san20}. Therefore, spatially resolved gas observations are critical in order to characterize the properties of gas and their connections to the star formation activity. 

To see if the SFE and \fh2~variations are driven by the adoption of a fixed $\alpha_{\mathrm{CO}}$, we repeated the same analyses with a metallicity and/or $\mathrm{CO}$ intensity-dependent $\alpha_{\mathrm{CO}}$ as described in \S \ref{sec:global}. While there is a systematic offset ($<$ 0.14 dex) in the resulted SFE and \fh2, the overall distributions of the SFE and \fh2~ remain very similar even if we adopt a varying $\alpha_{\mathrm{CO}}$. This suggests that the internal variations seen in the SFE and \fh2~ are intrinsic and not caused by the choices of $\alpha_{\mathrm{CO}}$. In a set of companion papers, we present in detail how the \fh2~and SFE are distributed spatially and which physical parameters are correlated most strongly with the dispersion of \fh2~and SFE (Ellison et al. 2020a,b; H.-A. Pan et al. in prep.).

\section{ALMaQUEST Key Science Papers}
The combination of spatially resolved stellar populations and emission line measurements from the MaNGA data and the gas properties derived from the ALMA observations enable a variety of kpc-scale investigations. While the primary goal of this paper is to present the main features of the ALMaQUEST survey, here we also highlight some key science applications, which are presented in more detail in a series of ALMaQUEST papers. 

\subsection{Kpc-scale scaling relations} \label{sec:rSFMS}
The process of star formation is described by two well-known scaling relations: the so-called `Schmidt-Kennicutt' relation \citep[SK;][]{sch59,ken98}, which relates the star formation rate to the underlying gas abundance, and the SFMS \citep{bri04,noe07,dad07,lin12,whi12,spe14}, a tight correlation between the star formation rate and the stellar mass. While the SK relation can be easily understood as the stars forming within molecular clouds, the origin of the SFMS has been hotly debated \citep[e.g.][]{kel14,tac16,hsi17,mat19,mor20,han20}. 

With the MaNGA-based measurements of \sigsfr~ and \sigsm~ and the ALMA-based \sigh2, we are able to discuss the relationships among the three quantities on kpc scales. In \citet{lin19b}, we showed that \sigsfr, \sigsm, and \sigh2~ computed for star-forming spaxels of MS galaxies form a linear 3D correlation in log scale. Each pair of these three quantities form a tight correlation, the resolved SFMS (rSFMS; i.e. \sigsfr~ vs. \sigsm), the resolved SK relation (rSK; i.e. \sigsfr~vs. \sigh2), and the resolved molecular gas main sequence (rMGMS; i. e. \sigh2~vs. \sigsm). By comparing the strength of the correlations and the magnitude of their scatters, we argue that rSFMS is the least fundamental, but rather a natural consequence of the combination of the rSK and rMGMS relations. This result is later supported by \citet{mor20}, who study the same scaling relations for five nearby spiral galaxies at 500pc scale, and by Ellison et al. (in prep.) based on a complementary approach applied to the ALMaQUEST sample.

%In \citet{lin19}, we present the relations for each pair of the three quantities and investigate which one(s) is more fundamental based on the rightness of the relations.

\subsection{What drives the central starburst?}
Previously, it has been shown that local starbursts are preferentially driven by central SFR enhancement \citep[e.g.,][]{mor17,ell18,tac18}. However, it remains unclear whether the boost of SFR in central regions is primarily caused by the increased gas fraction or greater star formation efficiency. Pinning down the relative importance between these two scenarios provides key constraints on the physical processes responsible for starbursts. Using the 12 starbursting galaxies from ALMaQUEST, we show in \citet{ell20a} that the central starburst is mainly driven by an elevated SFE. Only one quarter of the sample shows signs of mergers morphologically, indicating that other mechanisms may also produce central starbursts.

\subsection{What regulates the SFR in the star-forming main sequence?}
Recent IFS studies have shown that the tight correlation between global SFR and \sm~ may in fact originate from the rSFMS at kpc scales \citep[e.g.,]{can16,hsi17,ell18,pan18,can19}. As mentioned in \S. \ref{sec:rSFMS}, we further show that the rSFMS is due to a combination of both rSK relation and the rMGMS \citep{lin19b}. The rSFMS relation, however, shows a significant scatter $\sim$ 0.25 dex. In \citet{ell20b}, we investigate the dependence of the scatter around the rSFMS ($\Delta$\sigsfr) on the variation in \fh2~ and SFE to shed light on the physics regulating the scatter of star forming spaxels around the rSFMS. We found that the change in SFE is the primary cause for $\Delta$\sigsfr~ while the variation in \fh2~is a secondary factor \citep[but see also][]{dey19,mor20}.

\subsection{Gas properties in green valley galaxies}
Our pilot study \citep{lin17} of the CO content of three green valley galaxies suggests that the suppression of SFR in these `below MS' galaxies can possibly be attributed to the deficit in the gas fraction in the central regions of galaxies, accompanied with a reduction in both the gas fraction and SFE in the disks. With a sample size tenfold larger, we can more rigorously investigate the link between the star formation suppression and the changes in the gas fraction and SFE (L. Lin et al. in prep.). Also, we will characterize the radial distributions of gas fraction and SFE in green valley galaxies compared with those in main sequence galaxies (H.-A. Pan et al. in prep.).

\subsection{The non-universality of resolved scaling relations}
It is now well-established that global relations, such as the SK and SFMS, arise as a result of resolved scale correlations that exist on scales of kpc or less \citep{won02,big08,sch11,ler13,san13,can16,gon16,hsi17}.  The existence of such tight relations on kpc-scales indicates that they may reflect fundamental physical processes that are regulating the distribution of gas and its processing into stars.  Testing the variation of these scaling relations, and quantifying the global galactic properties on which they depend will therefore provide insight into the universality of the star formation process and its sensitivity to local environmental conditions.  Although past works have investigated the variation of either the rSK relation \citep[e.g.][]{sch11,ler13} or the rSFMS \citep[e.g.][]{abd17,pan18,vul19,can19}, the ALMaQUEST sample offers us the opportunity to investigate the variation in all three scaling relations (rSK, rSFMS and rMGMS), as well as study the interplay between them.  In Ellison et al. (in prep.) we demonstrate that all three of the resolved scaling relations exhibit significant variation, although this variation is significantly smaller for the rMGMS than for the rSFMS.  In Ellison et al. (in prep.) we also demonstrate that the variation in these scaling relations correlates with global galaxy properties such as total stellar mass, Sersic index and sSFR.

\subsection{The cold molecular gas-metallicity relation in MaNGA galaxies} 
%A large body of literature has discussed the relation between stellar mass, metallicity and SFR (or gas mass) for the local galaxy population \citep[e.g.,][]{ell08,man10,sal14}. The ALMaQUEST sample allows us to extend the study of this relation to local scales, and study to what extent, if at all, metallicity is affected by local changes in SFR and/or gas fraction.

Several works based on SDSS spectroscopy have demonstrated the existence of a three-dimensional relation between metallicity, stellar mass and SFR \citep[the so-called fundamental metallicity relation or FMR;][]{lar10, man10, sal14}. Such a relation is naturally predicted by several theoretical models \citep{dav11, lag16, tor19, tra19} as a consequence of a more fundamental link between metallicity, stellar mass and gas content. Models predict that the scatter across the mass-metallicity relation is driven by the competition between phases dominated by gas accretion and metal dilution and subsequent periods of enrichment and low gas fraction. In accordance with the observed FMR, at fixed stellar mass galaxies with higher SFR are predicted to have lower metallicities.

\cite{bot13} and \cite{bro18} studied the role of atomic gas using SDSS spectroscopy matched with H\textsc{i} observations from the  ALFALFA survey. Both studies agreed that the FMR is stronger when considering H\textsc{i} gas mass, rather than SFR as the parameter driving the scatter across the mass-metallicity relation. \cite{bot16a, bot16b} revise the role of gas on the FMR by considering molecular gas. They find that \Mh2~ is the best third parameters for the FMR, and is favoured over both total (atomic + molecular) gas mass or SFE. These findings provide observational confirmation of the importance of gas content, and in particular star-forming molecular gas, in driving the FMR.

Extensions of global scaling relations involving metallicity to resolved kpc-scale regions have been investigated thanks to large IFS surveys, like CALIFA, SAMI and MaNGA. A resolved mass-metallicity relation \citep{san13, bar16} is found to exist on kpc scales, and the existence of a secondary dependence of $\rm \Sigma_{SFR}$ is subject of active research (\citealt{bar16}; Belfiore et al. in prep. Also see Sec. 6.4 of \citealt{mai19} for a recent review). With ALMaQUEST we are now able to investigate the importance of  \sigh2~ in setting the scatter of the resolved mass-metallicity relation and assess its relative importance with respect to $ \rm \Sigma_{SFR}$.

\subsection{The connection between the Balmer Decrement (BD=\ha/\hb) and the CO(1-0) line luminosity  and total molecular gas mass}
The dust absorption, traced by the reddening and extinction of starlight has been used to estimate the gas content in the Milky Way \citep[e.g. see][]{boh78,lad94,pin10} and more recently in external galaxies \citep{bri13,con19,pio20}. In particular, \citet{con19} found an empirical relation between the dust extinction, traced by the Balmer Decrement (BD=\ha/\hb), and the total molecular gas mass in a sample of 222 local galaxies. By following this approach, we will explore whether the local BD-M$_{gas}$ relation and its connection with gas metallicity could be applied at local scales \citep[see also][]{bar20}, testing a new empirical method to trace the cold gas reservoir in galaxies. The ALMaQUEST sample allows us to extend the study of such relation in galaxies that are above, below and on the SFMS.

\section{SUMMARY}\label{sec:summary}

We introduce the ALMaQUEST survey, an ALMA program that maps the CO distributions on kpc scales for 46 galaxies selected from the MaNGA IFS survey. Whereas the MaNGA data deliver kpc-scale maps of star formation rate surface density, stellar mass surface density and metallicity, the ALMA observations provide spatially-matched maps of molecular gas.  Combined, the ALMA+MaNGA dataset yields a superlative view of star formation in nearby galaxies. The targets of the sample include starburst (SB), main sequence (MS) and green valley (GV) galaxies, allowing one to study the properties of cold ISM, star formation, stellar population, and ionized gas systematically across various galaxy populations. 

%When looking at the global(aperture-matched) stellar mass, \h2~mass, and star formation rate, it is found that the three galaxy populations (SB, MS, and GV) show distinct distributions in the Schmidt-Kennicutt relation. On the other hand, they largely overlap in the \h2~mass vs. stellar mass plane, although some GVs do exhibit significantly lower \h2~mass than SB and MS galaxies at a given stellar mass. In other words, the three subsamples are more segregated by their star formation efficiency (SFE) than their molecular gas fractions (\fh2). As the three subsamples are selected not only according to the specific star formation rate (sSFR) by also their detailed sSFR radial profiles (specifically for the SB population), the trends distinguishing the three classes may not be associated with the sSFR directly. In order to compare the relative contributions to the global sSFR between SFE and \fh2, we also study how the distributions of galaxies in the SK plane and the \h2~mass vs. stellar mass plane vary with sSFR. Through the correlation analysis, we find that sSFR has a stronger dependence on SFE than on \fh2. 

When looking at the global (aperture-matched) stellar mass, \h2~mass, and star formation rate, it is found that the locations of galaxies with respect to the Schmidt-Kennicutt relation (i.e., the SFE) is closely related  to the sSFR (see the middle panel of Figure \ref{fig:global_2d_ssfr} and the left panel of Figure \ref{fig:global_sfefgas_ssfr}). On the other hand, there exists a large scatter in the \h2~mass vs. stellar mass relation (and hence \fh2) at a given sSFR (see the bottom panel of Figure \ref{fig:global_2d_ssfr} and the right panel of Figure \ref{fig:global_sfefgas_ssfr}), although galaxies with low sSFR do exhibit significantly lower \h2~mass as opposed to those with high sSFR at a given stellar mass (bottom panel of Figure \ref{fig:global_2d_ssfr}). In general, galaxies with different sSFR are more segregated by their star formation efficiency (SFE) than their molecular gas fractions (\fh2). This is further supported through a correlation analysis, in which we find that sSFR has a slightly stronger dependence on SFE than on \fh2. On the other hand, we show that there is weaker dependence of sSFR on the atomic gas fraction (\MHI/\sm) or the molecular-to-atomic gas fraction (\Mh2/\MHI) compared to the dependence on SFE and/or \fh2. However, a larger and deeper \hi~ sample is required to draw a robust conclusion.

When comparing the CO and \ha~distributions within individual galaxies, we find that these two quantities do not always trace each other. We show that in a given galaxy, there is substantial variation in the sSFR, \fh2, and SFE for regions classified as star-forming spaxels (Figures \ref{fig:ssfr} -- \ref{fig:sfe}). Therefore, using a single global measurement may not be able to capture the detailed physics regulating the star formation within a galaxy. In the forthcoming papers, we will investigate in more detail the changes of the sSFR, SFE, and \fh2, as well as the correlations among these three parameters, as a function of various global and local galactic properties.

\acknowledgments

We thank the anonymous referee for his/her helpful comments. This work is supported by the Academia Sinica under
the Career Development Award CDA-107-M03 and the Ministry of Science \& Technology of Taiwan
under the grant MOST 107-2119-M-001-024 - and 108-2628-M-001 -001 -MY3. SFS thanks CONACYT CB-285080 and FC-2016-01-1916, and PAPIIT-DGAPA-IN100519 (UNAM) grants for supporting this project. RM acknowledges ERC Advanced Grant 695671 `QUENCH' and support by the Science and Technology Facilities Council (STFC). L. Lin and H.-A. Pan thank U. of Victoria for hosting during the visit to work on this project. L. Lin thanks Sophia Y. Dai and Z. Zheng for providing useful suggestions to improve the content of this paper.

The authors would like to thank the staffs of the East-Asia and North-America ALMA ARCs for their support and continuous efforts in helping produce high-quality data products. This paper makes use of the following ALMA data: ADS/JAO.ALMA\#2015.1.01225.S,  ADS/JAO.ALMA\#2017.1.01093.S, ADS/JAO.ALMA\#2018.1.00541.S, and ADS/JAO.ALMA\#2018.1.00558.S.
ALMA is a partnership of ESO (representing its member states), NSF (USA) and NINS (Japan), together with NRC (Canada), MOST and ASIAA (Taiwan), and KASI (Republic of Korea), in cooperation with the Republic of Chile. The Joint ALMA Observatory is operated by ESO, AUI/NRAO and NAOJ.

Funding for the Sloan Digital Sky Survey IV has been
provided by the Alfred P. Sloan Foundation, the U.S.
Department of Energy Office of Science, and the Participating Institutions. SDSS-IV acknowledges support
and resources from the Center for High-Performance
Computing at the University of Utah. The SDSS web
site is www.sdss.org. SDSS-IV is managed by the Astrophysical Research Consortium for the Participating
Institutions of the SDSS Collaboration including the
Brazilian Participation Group, the Carnegie Institution
for Science, Carnegie Mellon University, the Chilean
Participation Group, the French Participation Group,
Harvard-Smithsonian Center for Astrophysics, Instituto
de Astrof\'isica de Canarias, The Johns Hopkins University, Kavli Institute for the Physics and Mathematics of the Universe (IPMU) / University of Tokyo, Lawrence
Berkeley National Laboratory, Leibniz Institut f\"ur Astrophysik Potsdam (AIP), Max-Planck-Institut f\"ur Astronomie (MPIA Heidelberg), Max-Planck-Institut f\"ur
Astrophysik (MPA Garching), Max-Planck-Institut f\"ur
Extraterrestrische Physik (MPE), National Astronomical Observatory of China, New Mexico State University,
New York University, University of Notre Dame, Observat\'ario Nacional / MCTI, The Ohio State University,
Pennsylvania State University, Shanghai Astronomical
Observatory, United Kingdom Participation Group, Universidad Nacional Aut\'onoma de M\'exico, University of
Arizona, University of Colorado Boulder, University of
Oxford, University of Portsmouth, University of Utah,
University of Virginia, University of Washington, University of Wisconsin, Vanderbilt University, and Yale University. 

%\begin{deluxetable*}{lccclc}
\begin{deluxetable*}{lcccc}
\tabletypesize{\scriptsize}
\tablewidth{0pt}
\tablecaption{ALMaQUEST Targets and CO(1-0) Sensitivities \label{tab:sample}}
\tablehead{
    \colhead{Plate-IFU} &
    \colhead{RA} &
    \colhead{DEC} &
    \colhead{MaNGA redshift} &
%    \colhead{category} &
    \colhead{$\sigma_{CO}^{(a)}$}\\
&(deg)&(deg)&&(Jy beam$^{-1}$ \kms)}
\startdata
7815-12705 & 318.990448 & 9.543076 & 0.029550 & 0.0591 \\
 7977-3703 & 333.052032 & 12.205191 & 0.027817 & 0.0355 \\
 7977-3704 & 332.798737 & 11.800733 & 0.027245 & 0.0487 \\
 7977-9101 & 331.122894 & 12.442626 & 0.026562 & 0.0472 \\
 7977-12705 & 332.892853 & 11.795929 & 0.027236 & 0.0493 \\
 8077-6104 & 42.032784 & -0.752316 & 0.046014 & 0.0278 \\
 8077-9101 & 41.643112 & -0.843537 & 0.043226 & 0.0365 \\
 8078-6103 & 42.416542 & -0.069851 & 0.028593 & 0.0919 \\
 8078-12701 & 40.880466 & 0.306821 & 0.026977 & 0.0566 \\
 8081-3704 & 49.821442 & -0.969631 & 0.054004 & 0.0699 \\
 8081-6102 & 49.940136 & -0.077189 & 0.037189 & 0.0461 \\
 8081-9101 & 47.772182 & -0.546538 & 0.028460 & 0.0470 \\
 8081-9102 & 49.845692 & 0.823470 & 0.034069 & 0.0862 \\
 8081-12703 & 50.391369 & -0.178368 & 0.025583 & 0.0378 \\
 8082-6103 & 49.782173 & 0.955959 & 0.024157 & 0.0203 \\
 8082-12701 & 48.896458 & -1.016286 & 0.027026 & 0.0279 \\
 8082-12704 & 49.949562 & -0.221145 & 0.132144 & 0.0300 \\
 8083-6101 & 50.504082 & -1.053930 & 0.026766 & 0.0559 \\
 8083-9101 & 50.138412 & -0.339960 & 0.038470 & 0.1030 \\
 8083-12702 & 50.245415 & -0.367683 & 0.021040 & 0.0409 \\
 8084-3702 & 50.636642 & -0.001213 & 0.022061 & 0.0275 \\
 8084-6103 & 50.741676 & 0.054137 & 0.035927 & 0.0498 \\
 8084-12705 & 51.027115 & -1.057859 & 0.025446 & 0.0452 \\
 8086-9101 & 57.242985 & -0.521120 & 0.040035 & 0.0439 \\
 8155-6101 & 53.814114 & -1.228609 & 0.037403 & 0.2599 \\
 8155-6102 & 52.621368 & 0.752068 & 0.030814 & 0.0406 \\
 8156-3701 & 55.592297 & -0.583196 & 0.052726 & 0.0246 \\
 8241-3703 & 126.461189 & 18.166689 & 0.029113 & 0.0307 \\
 8241-3704 & 126.568909 & 17.362452 & 0.066173 & 0.0411 \\
 8450-6102 & 171.748840 & 21.141676 & 0.041996 & 0.0194 \\
 8615-3703 & 320.826416 & 1.254980 & 0.018452 & 0.0422 \\
 8615-9101 & 319.919739 & 0.120941 & 0.033459 & 0.0297 \\
 8615-12702 & 320.159454 & 1.047277 & 0.020947 & 0.0248 \\
 8616-6104 & 322.980530 & 0.213767 & 0.054257 & 0.0249 \\
 8616-9102 & 322.749451 & -0.000594 & 0.030386 & 0.0479 \\
 8616-12702 & 322.306061 & -0.294765 & 0.030831 & 0.0209 \\
 8618-9102 & 319.271454 & 9.972303 & 0.043337 & 0.0391 \\
 8623-6104 & 311.780975 & 0.300461 & 0.097041 & 0.0276 \\
 8623-12702 & 310.217072 & 0.652804 & 0.026910 & 0.0698 \\
 8655-3701 & 356.751831 & -0.447387 & 0.071489 & 0.0535 \\
 8655-9102 & 358.221924 & -0.382447 & 0.045050 & 0.0206 \\
 8655-12705 & 357.651733 & -1.128090 & 0.045568 & 0.0201 \\
 8728-3701 & 57.699028 & -7.028787 & 0.028327 & 0.0701 \\
 8950-12705 & 194.733139 & 27.833445 & 0.025277 & 0.0545 \\
 8952-6104 & 204.933975 & 27.776474 & 0.028433 & 0.0386 \\
 8952-12701 & 204.683838 & 26.328539 & 0.028563 & 0.0393

\enddata
\tablecomments{$^{(a)}$The 1$\sigma$ sensitivity of the integrated ALMA CO intensity maps (see \S \ref{sec:alma}), calculated using the spectral window shown as the yellow area in Figure \ref{fig:spectraA}.  }
%\tablecomments{The column 'categoty' denotes the three subclasses of the ALMaQUEST targets: main sequence(`MS'), green valley (`GV'), and starburst (`SB') galaxies.}
\end{deluxetable*}

%\begin{landscape}
\begin{deluxetable*}{lcccccccc}
%\begin{table}
%\begin{tabular}{lccccccccccccccc}
%\begin{tabular}{lc}
%\hline
\tabletypesize{\scriptsize}
\tablewidth{0pt}
\tablecaption{Properties of ALMaQUEST Galaxies Measured within 1.5\Re \label{tab:property_15re}}
\tablehead{
    \colhead{Plateifu} &
    \colhead{Area} &
    \colhead{log$_{10}$(\sm/\Msolar)} &
    \colhead{log$_{10}$($\frac{\rm SFR}{\rm M_{\odot} yr^{-1}}$)$^{(a)}$} &
    \colhead{$S_{CO(1-0)}$} &
   \colhead{log$_{10}$(\Mh2/\Msolar)$^{(b)}$} &
  \colhead{log$_{10}$($\frac{\rm SSFR}{\rm yr^{-1}}$)} &
    \colhead{log$_{10}$($\frac{\rm SFE}{\rm yr^{-1}}$)} &
    \colhead{log$_{10}$\fh2}\\
&(kpc$^{2}$)&&&(Jy \kms)       
    }

%Plateifu & A$^{(a)}$ &  log$_{10}$\sm$^{(a)}$ & log$_{10}$SFR$^{(a)}$ & $S_{CO} %$^{(a)}$ & $S_{CO}$$^{(a)}$ & \Mh2$^{(a)}$ & log$_{10}$sSFR$^{(a)}$ & log$_{10}$SFE$^{(a)}$ log$_{10}$\fh2$^{(a)}$ & log$_{10}$\sm$^{(b)}$ & log$_{10}$SFR$^{(b)}$  & $S_{CO}$$^{(b)}$ & \Mh2$^{(b)}$ & log$_{10}$sSFR$^{(b)}$ & log$_{10}$SFE$^{(b)}$ & %log$_{10}$\fh2$^{(b)}$ \\
%\hline

\startdata

7815-12705 & 102.58 & 10.77 & 0.80 & $31.76 \pm 0.19$ & $9.728 \pm 0.003$ & -9.96 & -8.93 & -1.04 \\
 7977-3703 & 45.27 & 10.43 & 0.39 & $8.86 \pm 0.04$ & $9.119 \pm 0.002$ & -10.04 & -8.72 & -1.31 \\
 7977-3704 & 27.34 & 10.45 & -0.59 & $2.96 \pm 0.02$ & $8.626 \pm 0.003$ & -11.03 & -9.21 & -1.82 \\
 7977-9101 & 68.88 & 11.20 & -0.14 & $2.80 \pm 0.01$ & $8.587 \pm 0.002$ & -11.33 & -8.72 & -2.61 \\
 7977-12705 & 124.02 & 11.02 & 0.66 & $18.62 \pm 0.16$ & $9.422 \pm 0.004$ & -10.35 & -8.76 & -1.59 \\
 8077-6104 & 294.49 & 10.75 & 0.61 & $8.07 \pm 0.03$ & $9.524 \pm 0.002$ & -10.14 & -8.91 & -1.23 \\
 8077-9101 & 52.70 & 10.42 & -0.08 & $2.17 \pm 0.02$ & $8.897 \pm 0.003$ & -10.50 & -8.97 & -1.52 \\
 8078-6103 & 69.77 & 10.80 & 0.51 & $23.16 \pm 0.12$ & $9.558 \pm 0.002$ & -10.29 & -9.05 & -1.24 \\
 8078-12701 & 180.47 & 11.12 & 0.37 & $29.18 \pm 0.15$ & $9.608 \pm 0.002$ & -10.75 & -9.23 & -1.52 \\
 8081-3704 & 89.82 & 10.33 & 0.91 & $2.73 \pm 0.01$ & $9.195 \pm 0.002$ & -9.42 & -8.28 & -1.13 \\
 8081-6102 & 93.51 & 10.95 & -1.52 & $1.99 \pm 0.02$ & $8.728 \pm 0.004$ & -12.46 & -10.24 & -2.22 \\
 8081-9101 & 78.58 & 10.73 & 0.41 & $18.89 \pm 0.15$ & $9.469 \pm 0.003$ & -10.32 & -9.06 & -1.27 \\
 8081-9102 & 67.68 & 10.81 & 0.17 & $9.78 \pm 0.08$ & $9.335 \pm 0.003$ & -10.64 & -9.16 & -1.48 \\
 8081-12703 & 67.46 & 10.51 & -0.98 & $7.92 \pm 0.09$ & $8.995 \pm 0.005$ & -11.49 & -9.98 & -1.52 \\
 8082-6103 & 47.42 & 10.20 & 0.35 & $7.47 \pm 0.02$ & $8.920 \pm 0.001$ & -9.85 & -8.57 & -1.28 \\
 8082-12701 & 150.88 & 10.53 & 0.07 & $6.00 \pm 0.02$ & $8.925 \pm 0.001$ & -10.46 & -8.85 & -1.61 \\
 8082-12704 & 1442.02 & 11.66 & -0.31 & $5.22 \pm 0.03$ & $10.271 \pm 0.002$ & -11.97 & -10.58 & -1.39 \\
 8083-6101 & 121.71 & 10.57 & -0.79 & $18.23 \pm 0.18$ & $9.400 \pm 0.004$ & -11.36 & -10.19 & -1.17 \\
 8083-9101 & 153.03 & 11.30 & 0.42 & $17.48 \pm 0.23$ & $9.702 \pm 0.006$ & -10.88 & -9.28 & -1.60 \\
 8083-12702 & 93.19 & 11.24 & 0.51 & $34.22 \pm 0.14$ & $9.468 \pm 0.002$ & -10.73 & -8.96 & -1.78 \\
 8084-3702 & 29.16 & 10.43 & 0.48 & $21.34 \pm 0.13$ & $9.294 \pm 0.003$ & -9.94 & -8.81 & -1.13 \\
 8084-6103 & 54.86 & 10.71 & -0.65 & $10.29 \pm 0.11$ & $9.412 \pm 0.005$ & -11.36 & -10.06 & -1.29 \\
 8084-12705 & 58.34 & 10.60 & -0.06 & $6.70 \pm 0.03$ & $8.919 \pm 0.002$ & -10.67 & -8.98 & -1.69 \\
 8086-9101 & 133.88 & 11.12 & -0.09 & $4.35 \pm 0.02$ & $9.133 \pm 0.002$ & -11.21 & -9.22 & -1.99 \\
 8155-6101 & 151.71 & 11.18 & -3.03 & $4.52 \pm 0.17$ & $9.087 \pm 0.017$ & -14.21 & -12.12 & -2.09 \\
 8155-6102 & 120.80 & 10.38 & 0.33 & $9.55 \pm 0.03$ & $9.243 \pm 0.001$ & -10.05 & -8.91 & -1.14 \\
 8156-3701 & 118.32 & 10.27 & 0.81 & $1.58 \pm 0.01$ & $8.934 \pm 0.002$ & -9.46 & -8.12 & -1.33 \\
 8241-3703 & 70.70 & 10.23 & 0.23 & $4.14 \pm 0.02$ & $8.830 \pm 0.002$ & -9.99 & -8.60 & -1.40 \\
 8241-3704 & 353.02 & 11.06 & 1.22 & $12.62 \pm 0.09$ & $10.039 \pm 0.003$ & -9.84 & -8.82 & -1.02 \\
 8450-6102 & 152.67 & 10.31 & 0.63 & $4.75 \pm 0.02$ & $9.213 \pm 0.002$ & -9.68 & -8.59 & -1.09 \\
 8615-3703 & 11.00 & 10.23 & 0.42 & $23.18 \pm 0.22$ & $9.175 \pm 0.004$ & -9.81 & -8.75 & -1.06 \\
 8615-9101 & 83.07 & 10.78 & 0.01 & $6.77 \pm 0.03$ & $9.166 \pm 0.002$ & -10.77 & -9.15 & -1.61 \\
 8615-12702 & 137.43 & 10.21 & 0.10 & $5.27 \pm 0.02$ & $8.645 \pm 0.001$ & -10.10 & -8.54 & -1.56 \\
 8616-6104 & 226.96 & 10.90 & 0.21 & $4.26 \pm 0.02$ & $9.392 \pm 0.002$ & -10.69 & -9.18 & -1.50 \\
 8616-9102 & 149.27 & 10.44 & 0.62 & $15.68 \pm 0.07$ & $9.446 \pm 0.002$ & -9.81 & -8.82 & -0.99 \\
 8616-12702 & 271.43 & 10.93 & -0.36 & $1.66 \pm 0.01$ & $8.483 \pm 0.002$ & -11.29 & -8.84 & -2.45 \\
 8618-9102 & 65.61 & 10.47 & 0.42 & $5.52 \pm 0.04$ & $9.306 \pm 0.003$ & -10.05 & -8.89 & -1.16 \\
 8623-6104 & 283.83 & 11.34 & 1.20 & $6.32 \pm 0.04$ & $10.077 \pm 0.003$ & -10.14 & -8.88 & -1.27 \\
 8623-12702 & 125.06 & 10.96 & 0.11 & $25.97 \pm 0.14$ & $9.561 \pm 0.002$ & -10.86 & -9.45 & -1.40 \\
 8655-3701 & 197.71 & 11.27 & 1.16 & $25.68 \pm 0.27$ & $10.416 \pm 0.005$ & -10.11 & -9.26 & -0.85 \\
 8655-9102 & 93.75 & 10.31 & 0.25 & $1.93 \pm 0.01$ & $8.884 \pm 0.002$ & -10.06 & -8.64 & -1.42 \\
 8655-12705 & 188.39 & 10.51 & -1.55 & $2.11 \pm 0.01$ & $8.931 \pm 0.003$ & -12.06 & -10.48 & -1.57 \\
 8728-3701 & 51.92 & 10.85 & -1.67 & $3.51 \pm 0.04$ & $8.730 \pm 0.005$ & -12.52 & -10.40 & -2.12 \\
 8950-12705 & 42.46 & 10.73 & -0.39 & $23.41 \pm 0.23$ & $9.460 \pm 0.004$ & -11.12 & -9.85 & -1.27 \\
 8952-6104 & 132.94 & 10.60 & 0.45 & $8.28 \pm 0.03$ & $9.111 \pm 0.001$ & -10.15 & -8.66 & -1.49 \\
 8952-12701 & 127.50 & 10.78 & -0.66 & $6.06 \pm 0.04$ & $8.974 \pm 0.003$ & -11.44 & -9.63 & -1.81

\enddata 
\tablecomments{$^{(a)}$ Only spaxels classified as star-forming are included. $^{(b)}$ The uncertainty listed here only refers to the measurement error, not yet including the uncertainty in the CO-to-\h2~conversion factor.  }
\end{deluxetable*}
%\end{tabular}
%\end{table}
%\end{landscape}

\newpage

%\begin{landscape}
\begin{deluxetable*}{cccccccc}
%\begin{table}
%\begin{tabular}{lccccccccccccccc}
%\begin{tabular}{lc}
%\hline
\tabletypesize{\scriptsize}
\tablewidth{0pt}
\tablecaption{Properties of ALMaQUEST Galaxies Measured within the MaNGA Bundle Coverage\label{tab:property_manga}}
\tablehead{
\colhead{Plateifu} &
    \colhead{log$_{10}$(\sm/\Msolar)} &
    \colhead{log$_{10}$($\frac{\rm SFR}{\rm M_{\odot} yr^{-1}}$)$^{(a)}$} &
    \colhead{$S_{CO(1-0)}$} &
   \colhead{log$_{10}$(\Mh2/\Msolar)$^{(b)}$} &
  \colhead{log$_{10}$($\frac{\rm SSFR}{\rm yr^{-1}}$)} &
    \colhead{log$_{10}$($\frac{\rm SFE}{\rm yr^{-1}}$)} &
    \colhead{log$_{10}$\fh2}\\
&&&(Jy \kms)    
    }

%Plateifu & A$^{(a)}$ &  log$_{10}$\sm$^{(a)}$ & log$_{10}$SFR$^{(a)}$ & $S_{CO} %$^{(a)}$ & $S_{CO}$$^{(a)}$ & \Mh2$^{(a)}$ & log$_{10}$sSFR$^{(a)}$ & log$_{10}$SFE$^{(a)}$ log$_{10}$\fh2$^{(a)}$ & log$_{10}$\sm$^{(b)}$ & log$_{10}$SFR$^{(b)}$  & $S_{CO}$$^{(b)}$ & \Mh2$^{(b)}$ & log$_{10}$sSFR$^{(b)}$ & log$_{10}$SFE$^{(b)}$ & %log$_{10}$\fh2$^{(b)}$ \\
%\hline

\startdata
7815-12705 & 10.82 & 0.84 & $31.23 \pm 0.19$ & $9.720 \pm 0.003$ & -9.98 & -8.88 & -1.10 \\
 7977-3703 & 10.50 & 0.42 & $9.90 \pm 0.04$ & $9.167 \pm 0.002$ & -10.08 & -8.74 & -1.34 \\
 7977-3704 & 10.54 & -0.42 & $3.82 \pm 0.02$ & $8.737 \pm 0.002$ & -10.95 & -9.15 & -1.80 \\
 7977-9101 & 11.27 & -0.09 & $4.50 \pm 0.02$ & $8.794 \pm 0.002$ & -11.36 & -8.88 & -2.47 \\
 7977-12705 & 11.05 & 0.70 & $18.27 \pm 0.16$ & $9.414 \pm 0.004$ & -10.35 & -8.71 & -1.64 \\
 8077-6104 & 10.78 & 0.69 & $8.20 \pm 0.03$ & $9.531 \pm 0.002$ & -10.09 & -8.84 & -1.25 \\
 8077-9101 & 10.55 & -0.02 & $2.23 \pm 0.02$ & $8.907 \pm 0.003$ & -10.57 & -8.93 & -1.65 \\
 8078-6103 & 10.88 & 0.65 & $26.91 \pm 0.12$ & $9.623 \pm 0.002$ & -10.23 & -8.97 & -1.26 \\
 8078-12701 & 11.15 & 0.42 & $26.35 \pm 0.15$ & $9.564 \pm 0.002$ & -10.73 & -9.15 & -1.58 \\
 8081-3704 & 10.54 & 1.07 & $3.18 \pm 0.01$ & $9.261 \pm 0.002$ & -9.47 & -8.19 & -1.28 \\
 8081-6102 & 11.03 & -1.20 & $1.97 \pm 0.02$ & $8.722 \pm 0.004$ & -12.22 & -9.92 & -2.30 \\
 8081-9101 & 10.82 & 0.42 & $18.10 \pm 0.15$ & $9.450 \pm 0.004$ & -10.40 & -9.03 & -1.37 \\
 8081-9102 & 10.89 & 0.26 & $8.73 \pm 0.08$ & $9.286 \pm 0.004$ & -10.63 & -9.03 & -1.60 \\
 8081-12703 & 10.55 & -0.98 & $6.89 \pm 0.09$ & $8.934 \pm 0.006$ & -11.53 & -9.91 & -1.62 \\
 8082-6103 & 10.30 & 0.41 & $7.82 \pm 0.02$ & $8.940 \pm 0.001$ & -9.90 & -8.53 & -1.36 \\
 8082-12701 & 10.59 & 0.16 & $5.73 \pm 0.02$ & $8.905 \pm 0.001$ & -10.43 & -8.75 & -1.68 \\
 8082-12704 & 11.76 & -0.30 & $4.63 \pm 0.03$ & $10.219 \pm 0.003$ & -12.06 & -10.52 & -1.54 \\
 8083-6101 & 10.58 & -0.80 & $15.43 \pm 0.18$ & $9.327 \pm 0.005$ & -11.38 & -10.12 & -1.26 \\
 8083-9101 & 11.37 & 0.50 & $15.67 \pm 0.23$ & $9.655 \pm 0.006$ & -10.87 & -9.15 & -1.71 \\
 8083-12702 & 11.30 & 0.74 & $42.09 \pm 0.14$ & $9.558 \pm 0.001$ & -10.56 & -8.82 & -1.74 \\
 8084-3702 & 10.46 & 0.50 & $21.79 \pm 0.13$ & $9.303 \pm 0.003$ & -9.96 & -8.80 & -1.16 \\
 8084-6103 & 10.79 & -0.65 & $10.01 \pm 0.11$ & $9.400 \pm 0.005$ & -11.44 & -10.05 & -1.39 \\
 8084-12705 & 10.65 & 0.00 & $5.83 \pm 0.04$ & $8.858 \pm 0.003$ & -10.65 & -8.86 & -1.79 \\
 8086-9101 & 11.22 & 0.08 & $5.19 \pm 0.02$ & $9.209 \pm 0.002$ & -11.14 & -9.13 & -2.01 \\
 8155-6101 & 11.22 & -3.03 & $4.84 \pm 0.19$ & $9.116 \pm 0.017$ & -14.26 & -12.15 & -2.11 \\
 8155-6102 & 10.43 & 0.40 & $10.38 \pm 0.03$ & $9.279 \pm 0.001$ & -10.03 & -8.88 & -1.15 \\
 8156-3701 & 10.41 & 0.91 & $1.94 \pm 0.01$ & $9.024 \pm 0.002$ & -9.50 & -8.11 & -1.38 \\
 8241-3703 & 10.26 & 0.27 & $4.23 \pm 0.02$ & $8.838 \pm 0.002$ & -9.99 & -8.57 & -1.42 \\
 8241-3704 & 11.09 & 1.25 & $12.58 \pm 0.09$ & $10.037 \pm 0.003$ & -9.83 & -8.78 & -1.05 \\
 8450-6102 & 10.39 & 0.66 & $4.96 \pm 0.02$ & $9.232 \pm 0.002$ & -9.72 & -8.57 & -1.15 \\
 8615-3703 & 10.34 & 0.44 & $29.80 \pm 0.23$ & $9.284 \pm 0.003$ & -9.90 & -8.84 & -1.06 \\
 8615-9101 & 10.84 & 0.05 & $6.13 \pm 0.03$ & $9.123 \pm 0.002$ & -10.79 & -9.08 & -1.72 \\
 8615-12702 & 10.23 & 0.09 & $4.32 \pm 0.02$ & $8.559 \pm 0.002$ & -10.14 & -8.47 & -1.67 \\
 8616-6104 & 10.96 & 0.23 & $4.14 \pm 0.02$ & $9.379 \pm 0.002$ & -10.73 & -9.15 & -1.58 \\
 8616-9102 & 10.47 & 0.68 & $15.04 \pm 0.07$ & $9.428 \pm 0.002$ & -9.79 & -8.75 & -1.04 \\
 8616-12702 & 10.97 & -0.34 & $1.43 \pm 0.01$ & $8.416 \pm 0.003$ & -11.31 & -8.75 & -2.56 \\
 8618-9102 & 10.57 & 0.46 & $5.18 \pm 0.04$ & $9.278 \pm 0.003$ & -10.11 & -8.82 & -1.29 \\
 8623-6104 & 11.53 & 1.29 & $7.12 \pm 0.04$ & $10.129 \pm 0.002$ & -10.24 & -8.84 & -1.40 \\
 8623-12702 & 11.00 & 0.11 & $22.30 \pm 0.14$ & $9.495 \pm 0.003$ & -10.89 & -9.39 & -1.51 \\
 8655-3701 & 11.33 & 1.16 & $26.53 \pm 0.27$ & $10.430 \pm 0.004$ & -10.17 & -9.27 & -0.90 \\
 8655-9102 & 10.43 & 0.27 & $1.70 \pm 0.01$ & $8.829 \pm 0.002$ & -10.15 & -8.55 & -1.60 \\
 8655-12705 & 10.55 & -1.54 & $1.51 \pm 0.01$ & $8.786 \pm 0.004$ & -12.09 & -10.33 & -1.76 \\
 8728-3701 & 10.88 & -1.67 & $3.34 \pm 0.04$ & $8.709 \pm 0.005$ & -12.55 & -10.38 & -2.17 \\
 8950-12705 & 11.16 & -0.38 & $22.77 \pm 0.23$ & $9.448 \pm 0.004$ & -11.54 & -9.83 & -1.72 \\
 8952-6104 & 10.62 & 0.46 & $8.40 \pm 0.03$ & $9.117 \pm 0.001$ & -10.16 & -8.65 & -1.51 \\
 8952-12701 & 10.86 & -0.66 & $5.25 \pm 0.04$ & $8.912 \pm 0.004$ & -11.52 & -9.57 & -1.94

\enddata
\tablecomments{$^{(a)}$ Only spaxels classified as star-forming are included. $^{(b)}$ The uncertainty listed here only refers to the measurement error, not yet including the uncertainty in the CO-to-\h2~conversion factor.}
\end{deluxetable*}

\addtocounter{figure}{+2}
\begin{figure*}
\centering
\includegraphics[angle=0,width=0.9\textwidth]{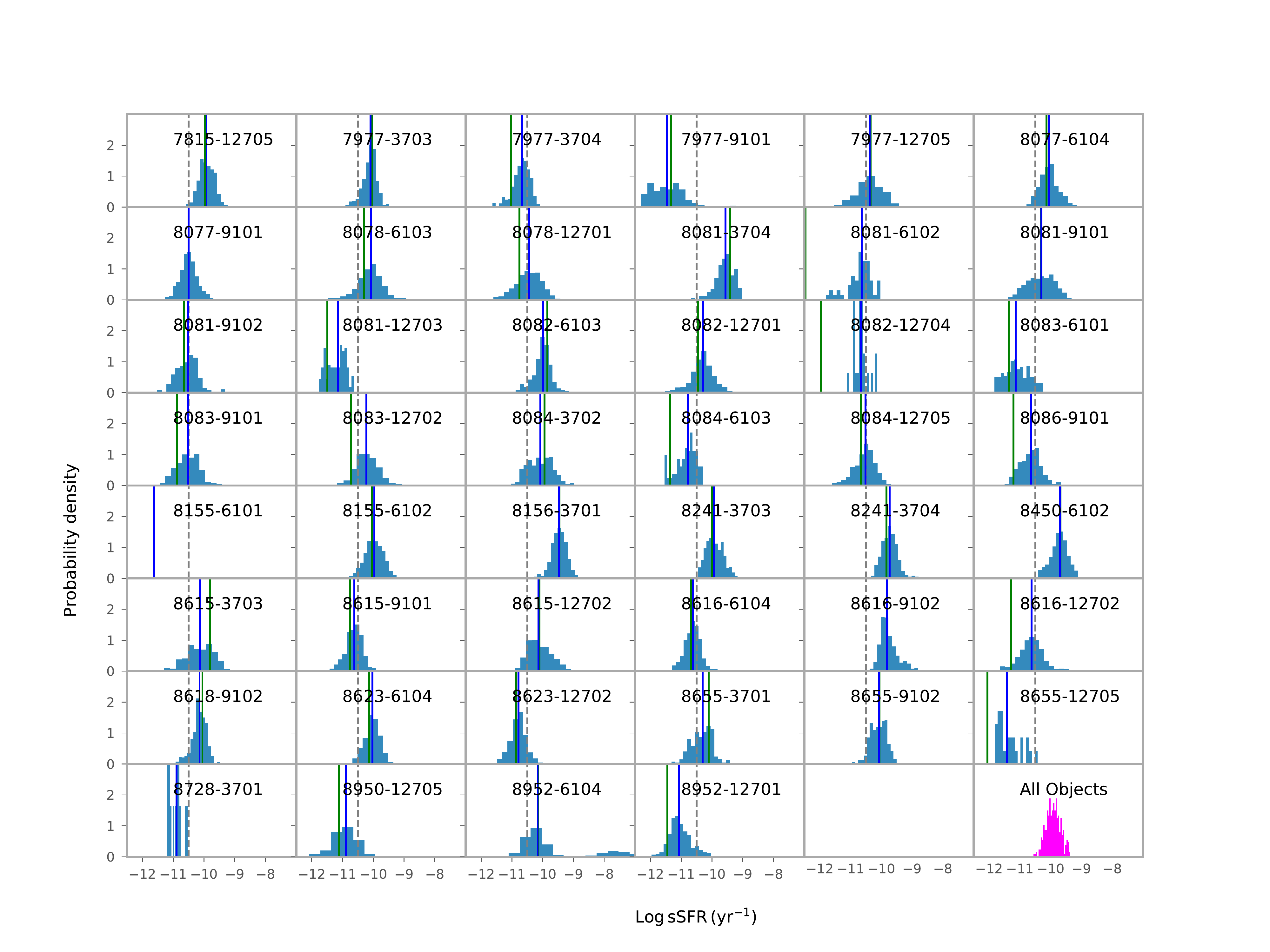}
\caption{Distributions of sSFR calculated on a spaxel-by-spaxel basis for 46 ALMaQUEST galaxies using only star-forming spaxels. The blue and green vertical lines represent the median of the histogram and the globally averaged value (within 1.5\Re) of individual galaxy, respectively. The gray dashed lines represent the constant value of Log$_{10}$sSFR (yr$^{-1}$) = -10.5 as a reference line. In some galaxies (e.g., 8082-12704), the global averaged value is very different to the median value of the histogram because the fraction of star-forming spaxels is low. The last panel shows the distribution of spaxels from all 46 galaxies. \label{fig:ssfr}}
\end{figure*}

\begin{figure*}
\centering
\includegraphics[angle=0,width=0.9\textwidth]{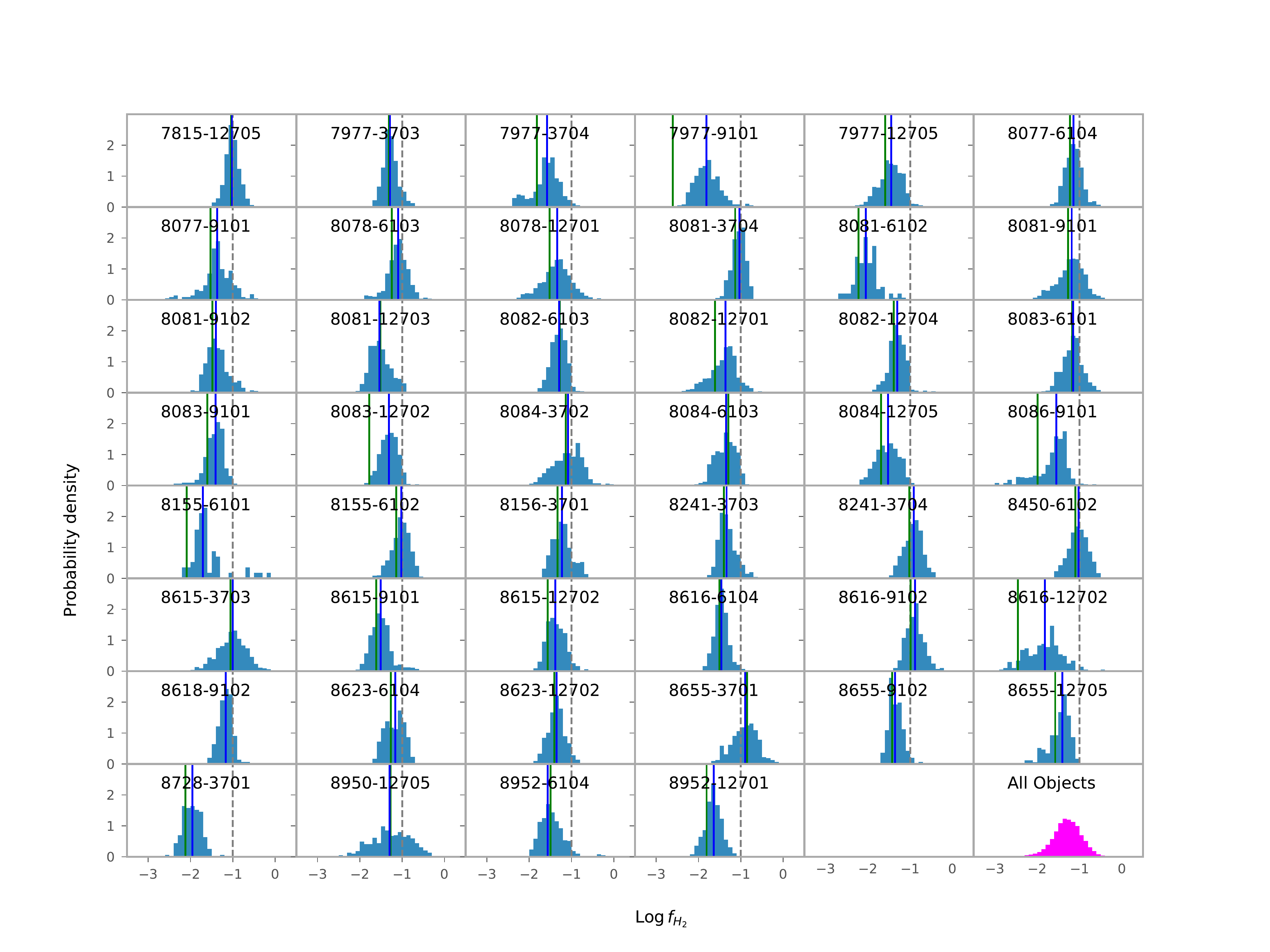}
\caption{Distributions of \fh2~calculated on a spaxel-by-spaxel basis for 46 ALMaQUEST galaxies. The blue and green vertical lines represent the median of the histogram and the globally averaged value (within 1.5\Re) of individual galaxy, respectively. The gray dashed lines represent the constant value of Log$_{10}$\fgas~ = -1 as a reference line. The histograms only represent spaxels with CO detections. In some galaxies (e.g., 8086-9101), the global averaged value is very different to the median value of the histogram because there is a significant fraction of spaxels without CO measurements. The last panel shows the distribution of spaxels from all 46 galaxies. \label{fig:fgas}}
\end{figure*}

\begin{figure*}
\centering
\includegraphics[angle=0,width=0.9\textwidth]{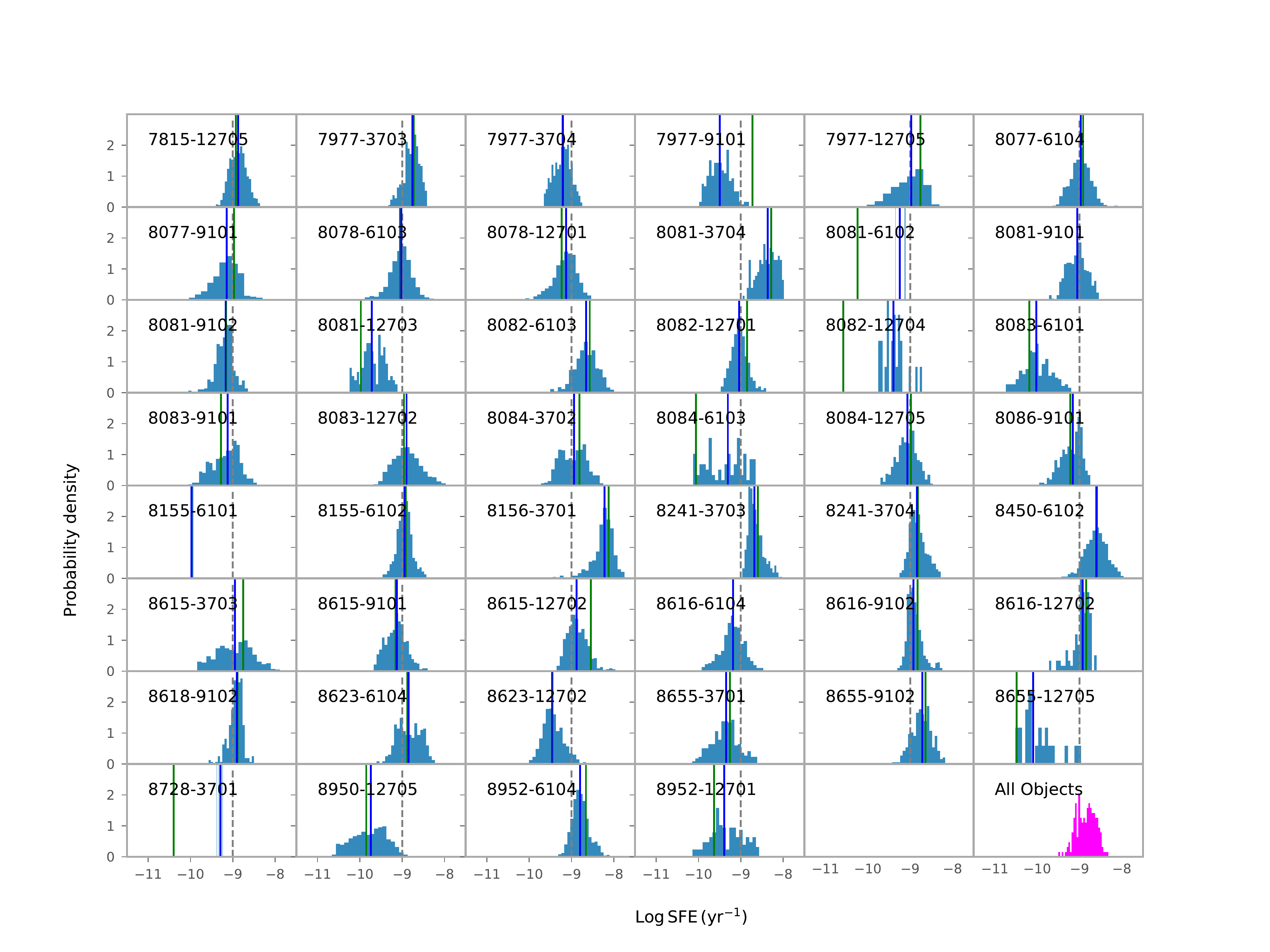}
\caption{Distributions of SFE~calculated on a spaxel-by-spaxel basis for 46 ALMaQUEST galaxies using only star-forming spaxels. The blue and green vertical lines represent the median of the histogram and the globally averaged value (within 1.5\Re) of individual galaxy, respectively. The gray dashed lines represent the constant value of Log$_{10}$SFE (yr$^{-1}$) = -9 as a reference line. The last panel shows the distribution of spaxels from all 46 galaxies. 
\label{fig:sfe}}
\end{figure*}

\appendix
\section{Tests on the effect of the global quantity estimates}

\numberwithin{equation}{section}
\renewcommand{\thefigure}{A\arabic{figure}}
\setcounter{figure}{0}

The main results concerning the global SFR of the ALMaQUEST sample presented in this work are primarily based on the SFR integrated over the areas classified as 'star-forming' using the [SII] BPT diagnostic \citep{kew01,kew06}. In this section, we test whether our results, specifically for the relation between sSFR and SFE and between sSFR and \fh2, are impacted by the choices of integrated areas or not. We repeat our analyses in two cases: a) by summing \sigsfr~over all spaxels within 1.5 \Re~regardless of the BPT types, and b) by summing \sigsfr~over only the star-forming spaxels (also within 1.5 \Re) classified by the [NII] BPT diagnostic \citep{kau03}, to be compared with the results presented in \S \ref{sec:global}, which is based on the [SII] BPT method. The caveat of the case (a) is that the SFR could be overestimated because the \ha~emissions might be powered by mechanisms other than star formation in non-star forming spaxels.

\begin{figure}
\centering
\plottwo{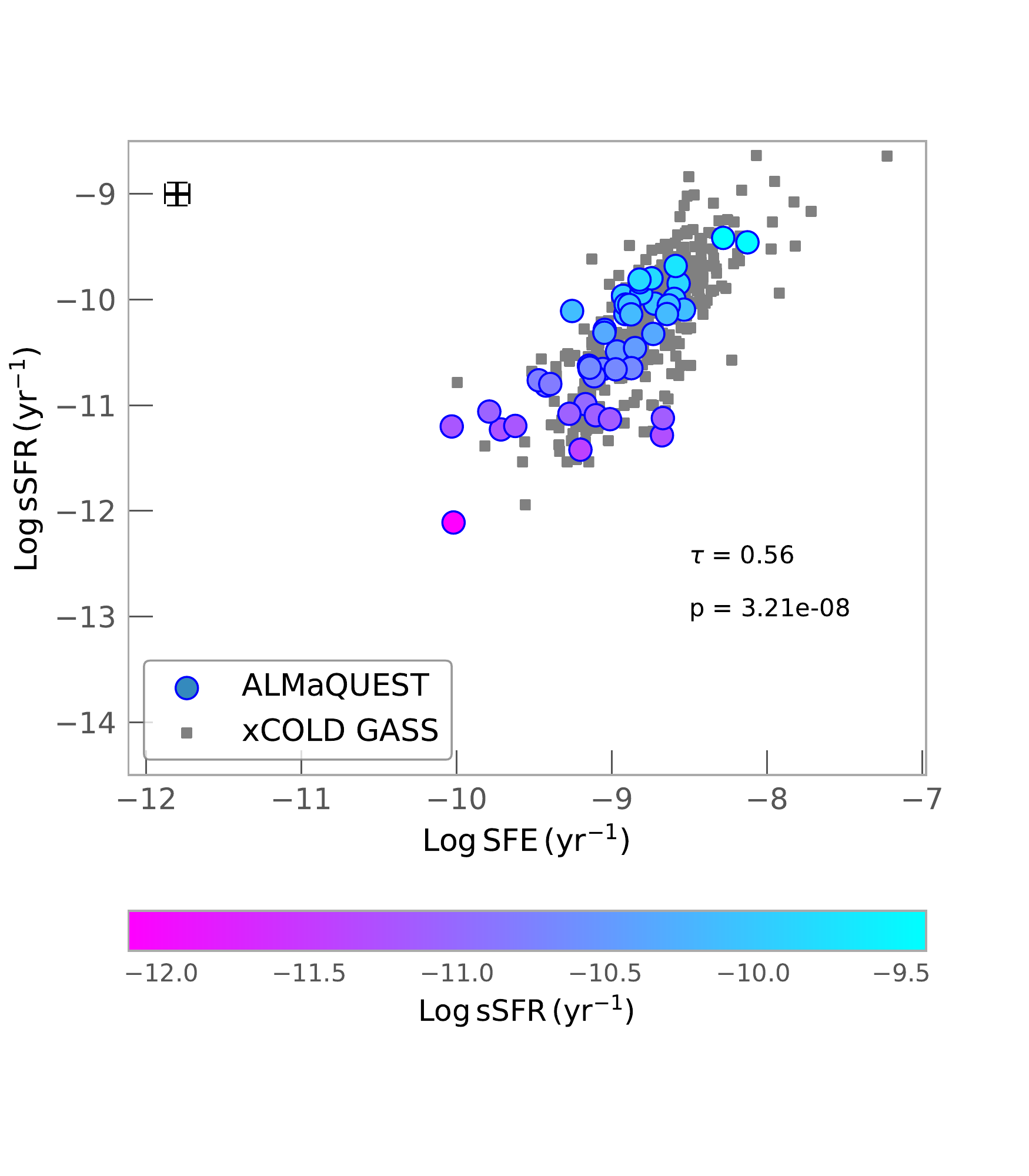}{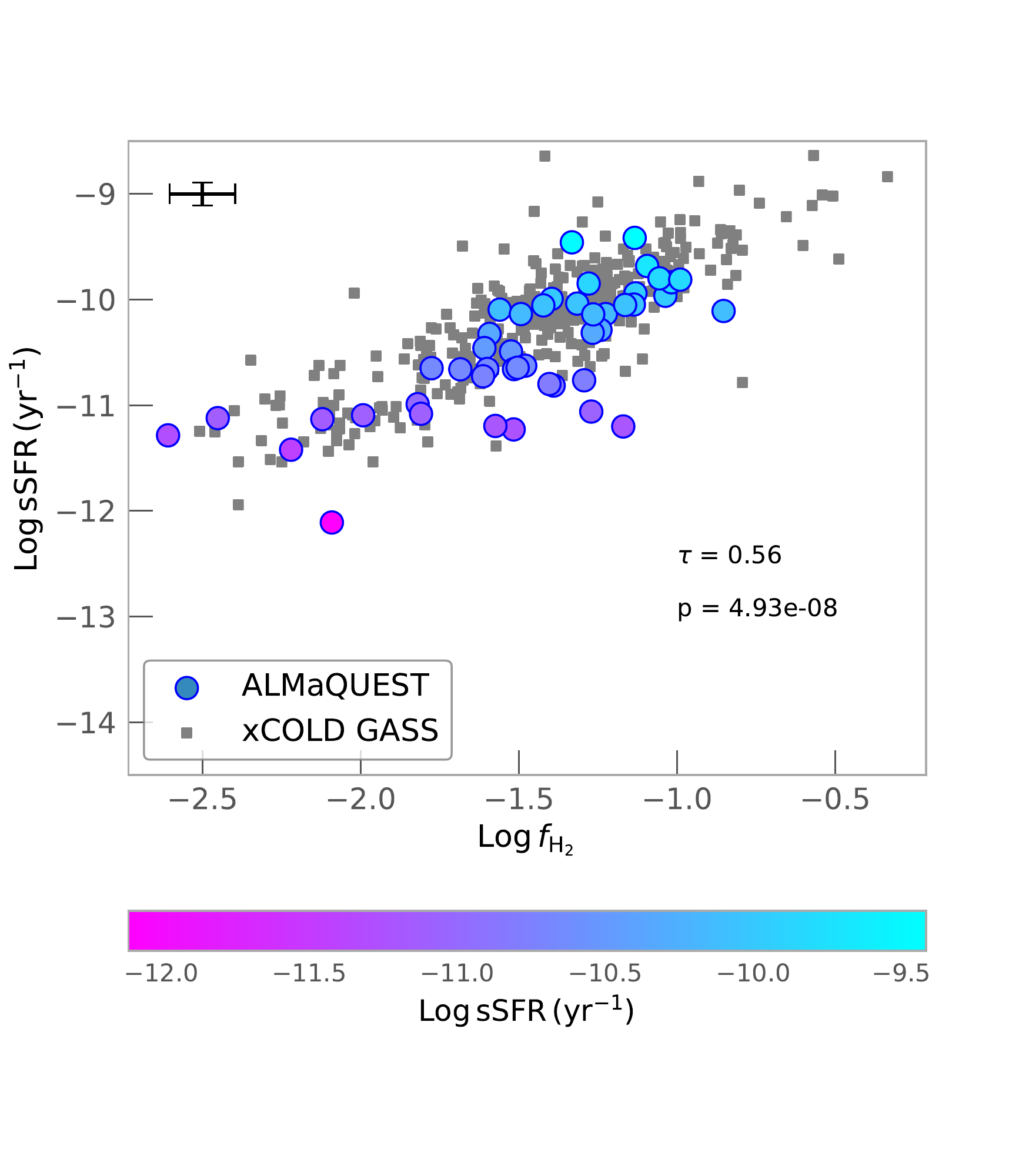}
\caption{Similar to Figure \ref{fig:global_sfefgas_ssfr} but with the SFR integrated using all spaxels regardless of the BPT line ratios.  \label{fig:global_sfefgas_ssfr_allspaxel}}
\end{figure}

\begin{figure}
\centering
\plottwo{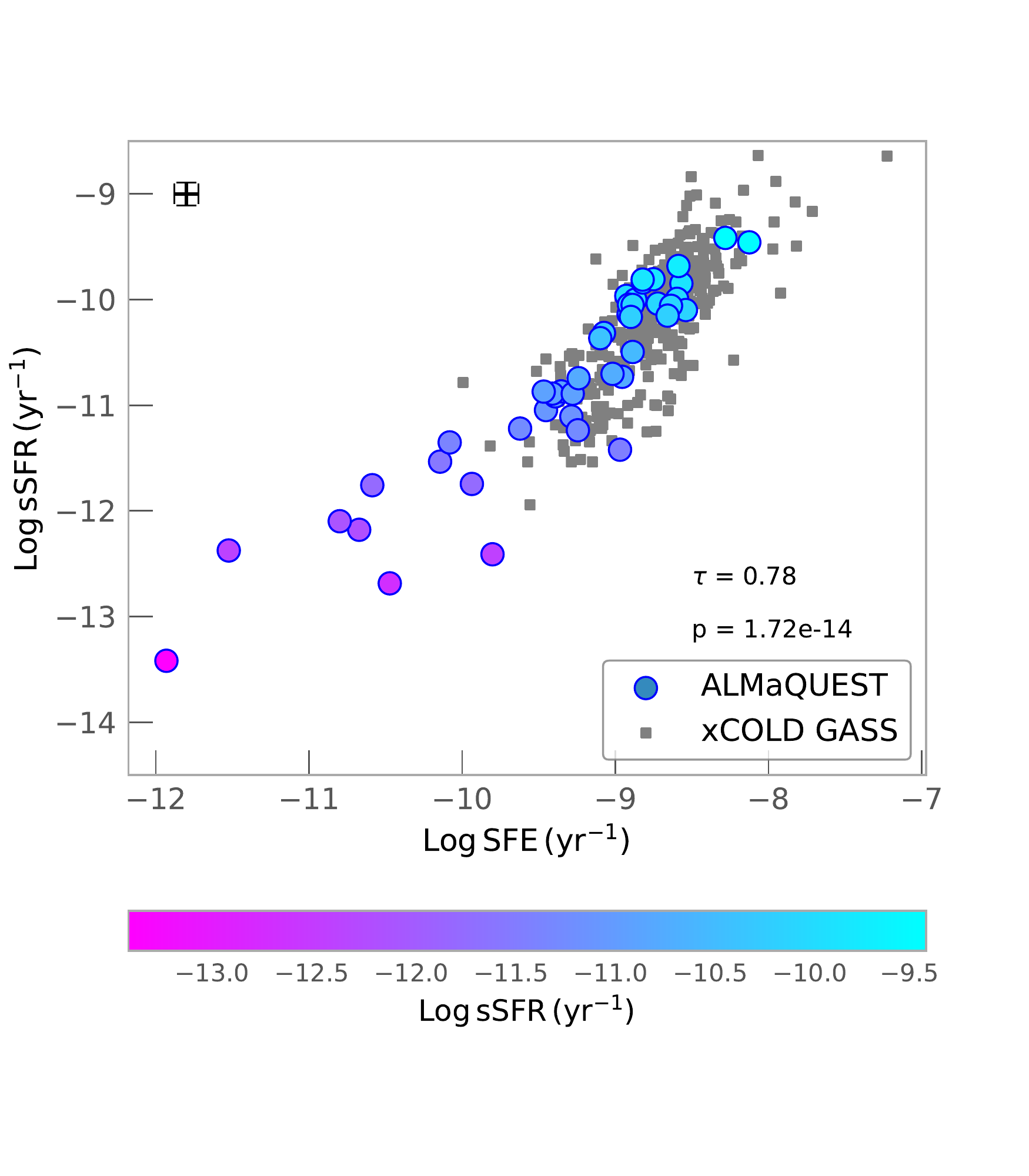}{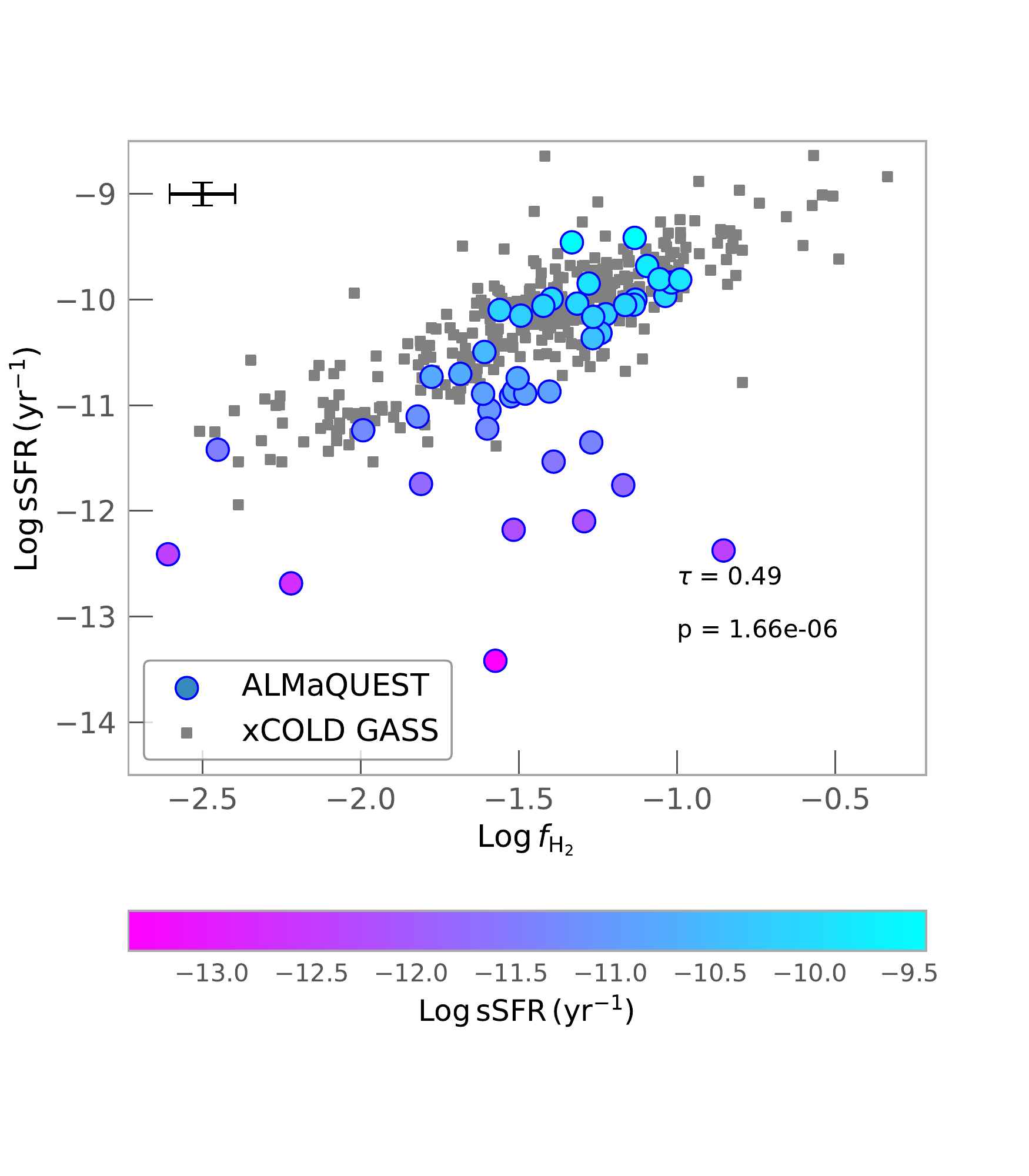}
\caption{Similar to Figure \ref{fig:global_sfefgas_ssfr} but with the SFR integrated over star-forming spaxels classified using  the [NII] BPT diagnostic \citep{kau03}.  \label{fig:global_sfefgas_ssfr_n2}}
\end{figure}

Figure \ref{fig:global_sfefgas_ssfr_allspaxel} displays the sSFR vs. SFE (left panel) and sSFR vs. \fh2~relations (right panel) when all types of spaxels are used. A moderate correlation ($\tau$ = 0.56) with high significance ($\rho < 10^{-7}$) is seen in both panels, suggesting that both SFE and \fh2~contribute to the variation of sSFR. In Figure \ref{fig:global_sfefgas_ssfr_n2} we plot the same relations by integrating only the [NII] BPT-classified star forming spaxels. Similar to Figure \ref{fig:global_sfefgas_ssfr}, a stronger correlation strength ($\tau$ = 0.78) is found in the sSFR vs. SFE relation, with respect to the sSFR vs. \fh2~relation ($\tau$ = 0.49). 

In summary, positive correlations are found in both the sSFR--SFE and sSFR--\fh2~relations in all the three situations we have tested, including the two cases presented here and the one in \S \ref{sec:global}. The strength of the correlation for the sSFR --SFE relation, is found to be more sensitive to the choices of the types of spaxels used for the total SFR calculation, whereas the sSFR--\fh2~ relation is more stable. Nevertheless, the main conclusion that the sSFR depends on both SFE and \fh2~hold in all cases.\\

\end{document}